%% file: complete_manuscript.tex
\documentclass[a4paper,fleqn,usenatbib,longnamesfirst]{mnras}
\usepackage[T1]{fontenc}
\usepackage{ae,aecompl}
\usepackage{graphicx}
\usepackage{amsmath}
\usepackage{amssymb}
\usepackage{booktabs}		
\usepackage{multirow}	
\usepackage{caption}		
\usepackage{subfig}		
\usepackage[separate-uncertainty=true]{siunitx}	
\usepackage{bm}	
\usepackage{lmodern}	
\usepackage{lipsum}
\usepackage{enumitem}
\usepackage[dvipsnames,usenames]{color}
\usepackage{hyperref}

\newcommand{\tup}[1]{\textup{#1}}
\newcommand{\msun}{$ \text{M}_\odot $}
\newcommand{\Mvir}{M_\tup{vir}}
\newcommand{\Rvir}{R_\tup{vir}}

\newcommand{\minus}{^{-1}}

\newcommand{\SZ}{Sunyaev--Zel'dovich}
\newcommand{\dg}{^{\circ}}
\DeclareSIUnit\kpc{kpc}
\definecolor{purple}{cmyk}{0,1.0,0,0.6}

\title[Kinetic SZ effect in synthetic rotating galaxy clusters]{Kinetic Sunyaev--Zel'dovich effect in rotating galaxy
	clusters from MUSIC simulations} %
\author[A.~S. Baldi et al.]{%
	Anna~Silvia Baldi,$^{1,2}$%
	\thanks{E-mail: \href{mailto:a.silvia.baldi@gmail.it}{\nolinkurl{annasilvia.baldi@uniroma1.it}}}
	Marco De Petris,$^{1}$
	Federico Sembolini,$^{1,3}$\newline
	\newauthor Gustavo Yepes,$^{3,4}$ Weiguang Cui,$^{3}$ Luca Lamagna$^{1}$
	\\
	\\
	$^{1}$Dipartimento di Fisica, Sapienza Universit\`{a} di Roma, Piazzale Aldo Moro 5, I-00185 Roma, Italy\\
	$^{2}$Dipartimento di Fisica, Universit\`{a} di Roma ``Tor Vergata'', Via della Ricerca Scientifica 1,
	I-00133 Roma, Italy\\
	$^{3}$Departamento de F\'{\i}sica Te\'{o}rica, M\'{o}dulo 8, Facultad de Ciencias, Universidad Aut\'{o}noma de
	Madrid, E-28049 Madrid, Spain\\
	$^{4}$Centro de Investigaci\'on Avanzada en F\'{\i}sica Fundamental (CIAFF), Universidad Aut\'onoma
	de Madrid, E-28049, Madrid, Spain
}

\date{Accepted XXX. Received YYY; in original form ZZZ}

\pubyear{2018}

\begin{document}
\graphicspath{{./}}
\label{firstpage}
\pagerange{\pageref{firstpage}--\pageref{lastpage}}
\maketitle

\defcitealias{cooray:ksz}{CC02}
\defcitealias{baldi:angmom}{B17}

\begin{abstract}
	The masses of galaxy clusters are a key tool to constrain cosmology through the physics of large-scale structure
	formation and accretion.
	Mass estimates based on X-ray and Sunyaev--Zel'dovich measurements have been found to be affected by the
	contribution of non-thermal pressure components, due e.g. to kinetic gas energy.
	The characterization of possible ordered motions (e.g. rotation) of the intra-cluster medium
	could be important to recover cluster masses accurately.
	We update the study of gas rotation in clusters through the maps of the kinetic Sunyaev--Zel'dovich effect,
	using a large sample of massive synthetic galaxy clusters ($ M_\tup{vir} > \num{5e14} h\minus$\msun\ at $z~=~0 $)
	from MUSIC high-resolution simulations. We select few relaxed objects showing peculiar rotational
	features, as outlined in a companion work.
	To verify whether it is possible to reconstruct the expected radial profile of the rotational velocity, we fit the
	maps to a theoretical model accounting for a specific rotational law, referred as the vp2b model.
	We find that our procedure allows to recover the parameters describing the gas rotational velocity profile within
	two standard deviations, both with and without accounting for the bulk velocity of the cluster.
	The amplitude of the temperature distortion produced by the rotation is consistent with theoretical
	estimates found in the literature, and it is of the order of 23 per cent of the maximum signal produced
	by the cluster bulk motion. We also recover the bulk velocity projected on the line of sight consistently
	with the simulation true value.
\end{abstract}


\input{./introduction}
\input{./music}
\input{./kszmaps}
\input{./results}
\input{./conclusions}

\section*{Acknowledgements}
This work has been supported by funding from Sapienza University of
Rome - Progetti di Ricerca Medi 2017, prot. RM11715C81C4AD67.
ASB acknowledges founding from Sapienza University of Rome - Progetti per Avvio alla Ricerca Anno 2017,
prot. AR11715C82402BC7.
WC and GY are supported by the Ministerio de Econom\'ia y Competitividad and the
Fondo Europeo de Desarrollo Regional (MINECO/FEDER, UE) in Spain through grant AYA2015-63810-P.
The authors thank P. Mazzotta for comments and suggestions, and acknowledge G. Cialone for useful discussions.
The MUSIC simulations were produced with the Marenostrum supercomputer at the
Barcelona Supercomputing Centre thanks to time awarded by Red Espa\~{n}ola de Supercomputaci\'{o}n.
This work makes an extensive use of \texttt{python}, particularly of the \texttt{scipy}
and \texttt{matplotlib} libraries.

\clearpage
\bibliographystyle{mnras}
\bibliography{./BIBLIOGRAPHY.bib}

\appendix
\input{./appendix}
\bsp	
\label{lastpage}
\end{document}

%% file: introduction.tex
\section{Introduction}
\label{sec:introduction}
The study of non-random motions within galaxy clusters is an important topic in modern astrophysics, especially to
determine accurate estimates of their masses.
Indeed, the most used methods to measure cluster masses are based on the
simple assumption of hydrostatic equilibrium \citep[see for instance][]{voit:clusterreview}, that accounts for the
pressure contribution due to random thermal motions only.
Results from numerical simulations of galaxy clusters show that
hydrostatic masses underestimate the real values by fractions of the order of $ ~ 10-20 $
per cent \citep[see e.g.][]{rasia:massbias06,meneghetti:massbias,nelson:massbias,biffi:hse}.
One of the possible explanations for this discrepancy is that it is necessary to account for 
additional pressure support coming both from turbulent motions \citep{rasia:ICMDMdensity,lau:motions},
and from coherent streams or rotation
\citep[as investigated by e.g.][]{fang:rotationandturbulence,biffi:velocity,biffi:phox}.
In the recent work by \citet{xcop:nonthermalpressure}, a quantitative estimate of the non-thermal pressure
support in the ICM is addressed, using for the first time observational X-ray and microwave data for a small sample
of nearby massive clusters.
The authors find that the median value of the non-thermal vs thermal pressure ratio is of the order of 10 per cent
at the virial radius, which is smaller than the predictions from numerical simulations.
Nevertheless, they also report an exception for the Abell cluster A2319, for which this ratio is
$\sim 50 $ per cent instead.

Ordered motions -- particularly rotation -- are, nevertheless, very challenging to measure.
As reviewed in \citet{hamden:clusterrotation}, several approaches for determining the presence of cluster rotation
exist. Indeed, it can be investigated from spectroscopic measurements towards the galaxy members at optical wavelengths
\citep{hwang:sdssrotating,tovmassian:galaxies,manolopoulou:clusterrotation},
or from observations of the diffuse intra-cluster medium (ICM hereafter) at X-ray
\citep{bianconi:articolo,liu:ccd} and microwave wavelengths \citep{cooray:ksz,chluba:ksz,sunyaev:turbulence}.
The latter, in particular, are possible through the \SZ\ (SZ) effect, that
is produced by the inverse Compton scattering of the hot free electrons in the ICM and the photons of the
cosmic microwave background (CMB) \citep{sz:1970,sz:1980}.
The SZ effect is observed as a variation in the temperature (or brightness) of the CMB spectrum,
and can be separated in two components, namely the thermal and the kinetic SZ effect
\citep[tSZ and kSZ, respectively; see e.g.][for a review]{rephaeli:sz}.
The tSZ is produced by the random thermal motion of the electrons, while the kSZ effect is attributed
to the motion of the gas in the cluster as a whole with respect to the CMB reference frame.
Quantitatively, the kSZ signal is proportional to the integral of the electron number density along the line of sight,
times the projection of the cluster velocity on the same line of sight.
This implies that, in principle, it would be possible to derive the kinetic properties of the ICM
from observations of the kSZ, as discussed e.g. in \citet{dupke:velocityfromksz}.
Some attempts to constrain cluster velocities with the kSZ have been reported in several works through
the years using data from simulations \citep[e.g. in][]{nagai:velocityfromksz}, as well as from observations
with different instruments with increasing sensitivity and angular resolution
\citep[see e.g.][]{holzapfel:ksz,benson:kszpeculiarvec,
	kashlinsky:kszstatistical,sayers:kszbolocam,sayers:kszlimit}.
For instance, a very recent example is the work by \citet{nika:tszandksz}, where they report the first high-significance
detection of the kSZ towards two cluster substructures in MACS J0717.5+3745 (with 3.4 and $ 5.1 \sigma $-significance,
respectively), from measurements at 260 GHz with an effective angular resolution of 22 arcseconds.
Also, cluster motions have been detected with high significance in a statistical fashion,
by estimating the pairwise momentum between couples of galaxy clusters using CMB data
\citep{hand:kszACT,planck:coppiette,chaoli:coppiette}.
It is worth to stress that, to quantify cluster velocities from single-cluster measurements, it is necessary to
use complementary data from X-ray observations, since the kSZ alone is not sufficient to disentangle the contribution
to the total signal coming from both the electron density -- that can be inferred from measurements of the
X-ray luminosity -- and the velocity.
Moreover, X-ray spectroscopy can also be an independent observational probe for the study of ICM motions.
For instance, recent observations of the Perseus cluster with the \textsl{Hitomi} satellite, allowed to
establish with high significance the presence of motions of the gas from measurements of the shift of
metal spectral lines with unprecedented resolution \citep{hitomi:2016,hitomi:2017}.

The possibility of exploiting the kSZ effect to detect rotational motions in clusters has been discussed in
\citet{cooray:ksz} \citepalias[][hereafter]{cooray:ksz} for the first time.
Assuming a simple solid body rotation for the gas, they
find that the kSZ temperature anisotropy induced by this motion would be characterized by a dipolar pattern,
produced by the projection on the line of sight of the velocity of the receding and approaching gas with
respect to the observer.
In \citet{chluba:ksz}, assuming the results of \citetalias{cooray:ksz} as a starting basis, they report a
detailed analytical development of the characteristics of this signal as a
function of cluster physical parameters, and they give some estimates of the expected amplitude, as measured with
interferometric methods towards a set of candidates from nearby clusters.
In both the aforementioned works, the amplitude of the signal at the dipole peak is estimated to range between few
\si{\micro\K} and tens of \si{\micro\K}, depending on the orientation of the line of sight with respect to the rotation
axis and on the dynamical state of the observed cluster.
Indeed, it is possible to explain the presence of a bulk rotation in galaxy clusters with the occurrence of a recent
merging event \citep[see e.g.][]{ricker:offaxismergers}, whose fetaures can be also inferred from the kSZ maps themselves,
as shown in the recent work of \citet{zhang:kszsimulated}.

At the time of writing no application of the kSZ for the study of cluster rotation has been reported so far.
In fact, this is a very challenging task which, observationally speaking, is primarily limited by the resolution reached
by current operating instruments, as well as from high relative errors of the inferred cluster velocity due to
instrumental and astrophysical contaminants, or to uncertainties in the reconstruction of complementary ICM properties
(e.g. the temperature).
Towards a possible future joint multi-wavelength detection of rotational motions in real clusters,
modern high-resolution gas-dynamical simulations are extremely valuable tools for preliminary analyses.
A very first test on the detectability of turbulent and ordered motions through the kSZ is reported in
\citet{sunyaev:turbulence}, where the authors consider a large, isolated galaxy cluster from a cosmological
simulation populated with dark matter and non-radiative gas physics. Its kSZ map shows the presence of a possible
coherent rotation, that can be deduced from two clearly distinguishable spots of opposite sign in the innermost
regions of the cluster.
In the present work, we take a step forward, making use of more realistic high-resolution cluster simulations
to improve the estimate of the rotational kSZ signal, and to possibly describe the ICM rotation with a more
suitable model for the rotational velocity, different from the simple solid body.
In particular, we study for the first time the temperature distortion
produced by the kSZ effect due to the ICM rotation, in massive galaxy clusters extracted from MUSIC simulations
\citep{sembolini:music1}.
We select a small sample of relaxed clusters showing also rotational features, as outlined
in a complementary work focused on the properties of the angular momentum and tangential velocity
of both ICM and dark matter in the same data set \citep[][B17 hereafter]{baldi:angmom}.
By applying \citetalias{baldi:angmom}'s vp2b model to describe the radial profile of the tangential velocity,
we test whether the expected rotational properties can be recovered from a fit to the kSZ maps, both with
and without accounting for the overall cluster bulk motion.
Since this preliminary analysis is not linked to any particular experiment, in this work we use clean data
without accounting for noise or contaminations of astrophysical and instrumental origin. The full treatment
of these effects will be included in a forthcoming work.

This paper is organized as follows.
In section~\ref{sec:music} we present the cluster data set used for this study. In section~\ref{sec:kszmaps}
we describe the theoretical and the synthetic kSZ maps, whose analysis is reported and discussed in 
section~\ref{sec:results}. We eventually summarize our conclusions in section~\ref{sec:conclusions}.

%% file: music.tex
\section{Data set}
\label{sec:music}
The simulated galaxy clusters analysed in this work constitute a small sub-sample of objects extracted from the
MUSIC\footnote{\texttt{\url{http://music.ft.uam.es}}} hydrodynamical $ N $-body simulation project
\citep[see][for details]{sembolini:music1}.
MUSIC simulations have been run with the \textsc{Gadget-3} TreePM+SPH code \citep{springel:gadget}.
The initial conditions were extracted from two different parent simulations:
MareNostrum Universe \citep{yepes:marenostrum}, constituting the MUSIC-1 sub-set, and MultiDark
\citep{prada:multidark}, constituting the MUSIC-2 sub-set.
All the clusters were simulated using two different models to describe the baryonic component: one including only
smoothed particle hydrodynamics (SPH) and gravity forces (NR subset), and one adding multi-phase inter-stellar medium
modelling and radiative physics processes including radiative cooling, UV photoionization, star formation and
supernova feedback (CSF subset), see \citet{sembolini:music1} for further details.
Here we focus on the 258 most massive objects from MUSIC-2, having virial mass $ \Mvir>\num{5e14}
h\minus$\msun\ at redshift $ z=0 $, that have been already analysed in previous works
to study SZ and X-ray scaling relations \citep{sembolini:music1,biffi:music2,sembolini:music3},
rotational features \citepalias{baldi:angmom} and SZ-derived morphological properties \citep{cialone:szmorphology}.

To avoid possible contamination from the motion of large substructures in the kSZ maps, we limit our analysis to a small
sub-sample of objects which feature (\emph{i}) a relaxed dynamical state, and (\emph{ii}) a sufficiently large 
spin parameter of the gas.
Following the same procedure of \citetalias{baldi:angmom}, we first determine the dynamical state using two indicators
derived from the simulation.
The first one is the offset between the position of the centre of mass of the cluster and the position of the
density peak, normalized to the virial radius, $ \Delta r $.
The second one is the ratio between the mass of the largest sub-structure in the cluster and
the virial mass of the cluster itself, $ M_\tup{sub}/\Mvir $.
The detailed relation between these two indicators and their connection with the estimator of the cluster dynamical
equilibrium can be found in \citet{cui:dynamicalstate}.
We select the clusters that fulfil both conditions $ \Delta r < 0.1 $ and $ M_\tup{sub}/\Mvir < 0.1 $.
After selecting these relaxed clusters, we apply a further selection based on the values of the spin parameter of the gas,
$ \lambda_\tup{gas} $, considering only those having $ \lambda_\tup{gas} > 0.07 $, as we did in \citetalias{baldi:angmom}.
With both these selection criteria, our final sample is reduced to six objects with possibly detectable rotation, whose main
properties are listed in Table~\ref{tab:clusterdata}.
The different modelling used to describe the baryon physics in MUSIC simulations (e.g. cooling and star formation)
has been found to be not relevant for the gas rotational properties \citepalias[see][]{baldi:angmom},
which is in agreement with the results of \citet{cui:nifty}. For this reason, we limit here our analysis to the
data from the non-radiative run, that does not account for radiative gas physics.
In these simulations the gas mass particle is set to $ m_\tup{gas} = \num{1.9e8} h\minus$\msun,
while dark matter particles have mass $ m_\tup{DM} = \num{9.0e8} h\minus$\msun.

We assume throughout the paper the same cosmological model adopted in MUSIC-2 and in the MultiDark parent simulation,
which takes the parameter values from the best fit to the \textsl{WMAP7}+BAO+SNI data: $ \Omega_m=0.27$,
$ \Omega_b=0.0469$, $ \Omega_{\Lambda}=0.73$, $ \sigma_8=0.82$, $ n=0.95$ and $ h=0.7 $
\citep{komatsu:wmap}.
\begin{table}
	\centering
	\caption{Identifier, virial mass, virial radius and spin parameter of the gas of
		the six relaxed and rotating MUSIC clusters analysed in this work (see text).}
	\begin{tabular}{cccc}
		\toprule
		cluster ID & $\Mvir$ ($ \times \num{e15} $ \msun) & $\Rvir$ (kpc) & $ \lambda_\tup{gas} $\\
		\midrule
		46 & 1.17 & 2756 & 0.0785 \\
		93 & 1.90 & 3241 & 0.0769 \\
		98 & 1.61 & 3071 & 0.0735 \\
		103 & 1.02 & 2633 & 0.0746 \\
		205 & 1.24 & 2813 & 0.0763 \\
		256 & 1.31 & 2867 & 0.0714 \\
		\bottomrule
	\end{tabular}
	\label{tab:clusterdata}
\end{table}

%% file: kszmaps.tex
\section{Methods}
\label{sec:kszmaps}
This section illustrates the theoretical model used to describe the rotation of the ICM in our clusters, and
the method we use to compute the kSZ maps from the available data.

\subsection{Theoretical kSZ maps}
\label{subsec:theoreticalmaps}
In the non-relativistic regime \citep{kompaneets:approx} the temperature shift produced by the kSZ
observed along the direction identified by the vector $ \hat n $, can be written as \citep[e.g.][]{rephaeli:sz}:
\begin{equation}
	\label{eqn:deltaT_ksz}
	\frac{\Delta T_\tup{kSZ}(\hat n)}{T_\tup{CMB}} = - \frac{\sigma_T}{c} \int\limits_\text{los} n_e v_p \ dl \ ,
\end{equation}
where $ T_\tup{CMB} $ is the CMB monopole temperature, $ T_\tup{CMB} = (2.725 \pm 0.001)$ K 
\citep{mather:tcmb,fixsen:tcmb} , $ \sigma_T $ is Thomson cross section,
$ c $ is the speed of light, $ n_e $ is the electron number density, and $ v_p $ is the projection of the gas
velocity on the line of sight (los).

\begin{figure}
	\centering
	\includegraphics[width=0.46\textwidth]{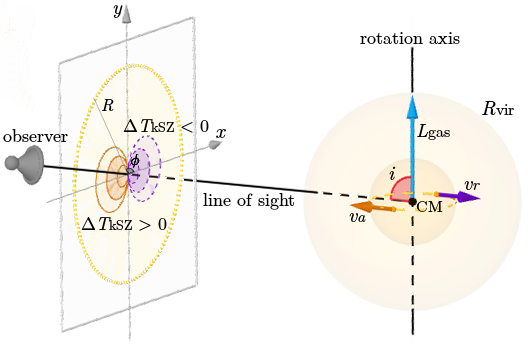}
	\caption{\small Sketch of the expected kSZ map from a rotating cluster, assuming the best observational
		configuration (i.e. with the line of sight orthogonal to the axis of rotation).
		The gas distribution in the cluster is assumed to be spherically symmetric, with the rotation axis aligned with
		the angular momentum vector of the gas as measured at the virial radius ($ \bm L_\tup{gas} $).
		The velocity vectors, $ \bm v_a $ and $ \bm v_r $, indicate respectively the approaching and receding
		velocity components along the line of sight for two generic gas particles, located at the same radial
		distance from the cluster centre of mass.}
	\label{fig:mapsketch}
\end{figure}

If the gas rotates according to a law described by a generic angular velocity $ \omega(r) $ along the
cluster radius, equation~\eqref{eqn:deltaT_ksz} for such rotational component becomes:
\begin{equation}
\label{eqn:coorayteo}
		\frac{\Delta T_\tup{kSZ,r}(R,\phi)}{T_\tup{CMB}} =
		- \frac{\sigma_T}{c} R \cos \phi \sin i %
		\int\limits_{R}^{R_\tup{vir}} n_e(r) \ \omega(r) \
		\frac{2 r \ dr}{\sqrt{r^2 - R^2}}
\end{equation}
as prescribed in \citetalias{cooray:ksz}.
$ R $ and $ \phi $ in equation~\eqref{eqn:coorayteo}
are the polar coordinates of the map in the plane of the sky, while $ r $ is the three-dimensional radial distance
from the cluster centre of mass.
The $ \sin i $ factor accounts for the orientation, given by the angle $ i $, of the line of sight with respect to the
rotation axis of the gas. If the line of sight is orthogonal or parallel with respect to
the rotation axis, the amplitude of the signal will be therefore maximum or null, respectively.
A simple sketch illustrating the best observational configuration (with $ i = 90\dg $) is represented in 
Fig.~\ref{fig:mapsketch}, in which the expected dipole-shaped pattern can be seen in the map as two symmetric
spots -- a positive one for approaching gas and a negative one for receding gas -- aligned horizontally with respect to
the projected axis of rotation.

In the most general case, the kSZ signal from a cluster is not only due to a pure rotational motion.
Indeed, a contribution from the cluster bulk velocity is also present, and it may be dominant with respect to the
rotation, depending on the projection of this velocity on the line of sight.
The full model describing the theoretical kSZ due to a possible rotation plus the bulk motion is therefore:
\begin{equation}
	\label{eqn:coorayteobulk}
	\begin{split}
		\frac{\Delta T_\tup{kSZ}(R,\phi)}{T_\tup{CMB}} &= \frac{\Delta T_\tup{kSZ,r}(R,\phi)}{T_\tup{CMB}} \ + \\
		&- \frac{\sigma_T}{c} \ v_\tup{bulk} %
		\int\limits_{R}^{R_\tup{vir}} n_e(r) \
		\frac{2 r \ dr}{\sqrt{r^2 - R^2}} \ .
	\end{split}
\end{equation}
We denote with $ v_\tup{bulk} $ the projection of the cluster bulk velocity of the gas on the line of sight;
the $ \Delta T_\tup{kSZ,r}(R,\phi)/T_\tup{CMB} $ term for the rotational component is given in
equation~\eqref{eqn:coorayteo}.
In this case, the expected signal is characterized by an asymmetric dipolar pattern, depending on the dominating
approaching or receding $ v_\tup{bulk} $ at a given line of sight.

As can be seen from equations~\eqref{eqn:coorayteo} and~\eqref{eqn:coorayteobulk}, the calculation of the theoretical
kSZ maps requires to compute the integral over the line of sight of the electron number density profile,
$ n_e(r) $, and of the angular velocity profile, $ \omega(r) $. The two analytical expressions we adopt to describe
these profiles are detailed as follows.
\begin{itemize}
	\item \emph{Electron number density}:
	in real-life observations of clusters, the SZ effect alone cannot be used to constrain all the thermodynamic
	and dynamical properties of the ICM, as pointed out also in section~\ref{sec:introduction}.
	Thus, in order to derive the parameters characterizing the rotational velocity it is necessary to
	have an independent measurement of the electron number density, and a possible estimate of the analytical
	model describing its radial profile, $ n_e(r) $.
	This information has to be provided by ancillary X-ray observations, that allow to recover cluster densities
	at radii up to the virial radius \citep[see e.g.][and references therein]{tchernin:a2142}.
	In our case, instead of using mock X-ray data, we make use of the numerical profiles derived from the cluster data
	provided by the simulation, and we fit them to a suitable theoretical model.
	The numerical profiles are computed as described in \citet{sembolini:music1}, following:
	\begin{equation}
	\label{eqn:nesim}
	n_e(r) = N_e(r) \ \rho_\tup{gas}(r) \ \frac{Y_H}{m_p} \ ,
	\end{equation}
	where $ N_e(r) $ and $ \rho_\tup{gas}(r) $ are the number of electrons and the gas density
	at the cluster radius $ r $, $ Y_H = 0.76$ is the hydrogen abundance referred to the gas particle,
	and $ m_p $ is the proton mass.
	The model we use to describe $ n_e(r) $ is a simplified six-parameter version of the equation proposed in
	\citet{vikhlinin:gasdensity}, that is:
	\begin{equation}
	\label{eqn:vikhlinin06}
	n_e(r) = n_0 \ \frac{(r/r_c)^{-\frac{\alpha}{2}}}{[1+(r/r_c)^2]^{\frac{3 \beta}{2} - \frac{\alpha}{4}}} \ 
	\frac{1}{[1+(r/r_s)^{\gamma}]^{\frac{\varepsilon}{2 \gamma}}} \ ,
	\end{equation}
	where $ n_0 $ is the central density, $ r_c $ and $ r_s $ are scale radii, and $ \alpha $, $ \beta $
	and $ \varepsilon $ control the slopes of the profile at different radii.
	To determine the best-fit values of the free parameters ($ n_0 $, $ r_c $, $ \alpha $, $ \beta $, $ r_s $ and
	$ \varepsilon $), we use a Markov chain Monte Carlo procedure. The slopes $ \alpha, \beta $ and $ \varepsilon $,
	are all constrained to be positive, with the additional condition $ \varepsilon < 5 $;
	the $ \gamma $ parameter instead is kept fixed to 3.0 \citep{vikhlinin:gasdensity}.
	We summarize the results in Fig.~\ref{fig:nefit}, showing the data and the best-fit curves, and in
	Table~\ref{tab:nefitparams}, where we list the parameter values.
	\begin{figure}
		\centering
		\includegraphics[width=0.48\textwidth]{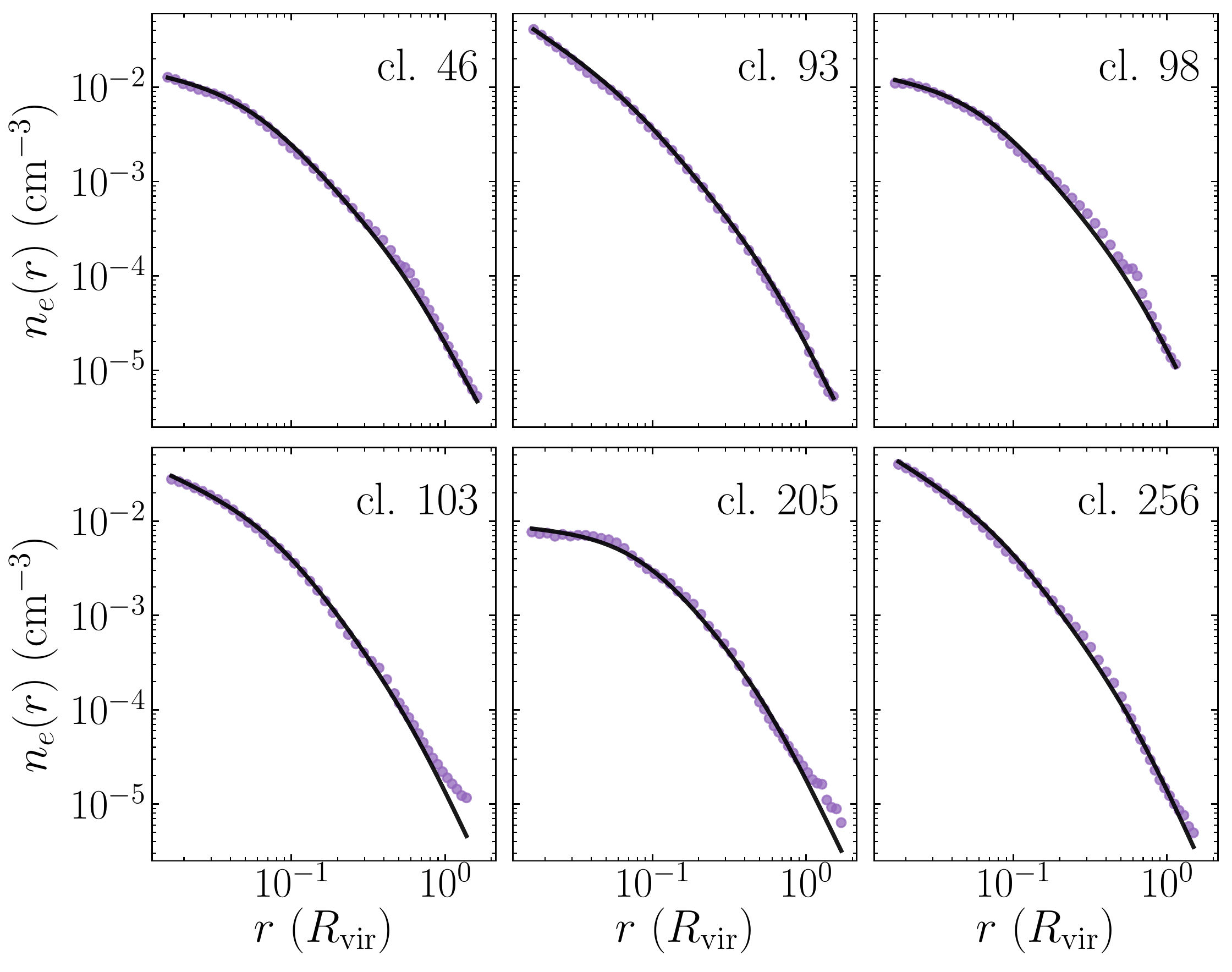}
		\caption{\small Radial profiles of the electron number density of the clusters in our sample.
			Purple dots are the median values computed according to equation~\eqref{eqn:nesim};
			solid black lines represent the best-fit curves described by the simplified Vikhlinin model of
			equation~\eqref{eqn:vikhlinin06} with the parameters listed in Table~\ref{tab:nefitparams}.}
		\label{fig:nefit}
	\end{figure}
	\begin{table*}
		\centering
		\caption{\small Parameter values from the fit of radial profiles of the electron number density derived from
			the simulation to the simplified Vikhlinin model of equation~\eqref{eqn:vikhlinin06}. The $ \gamma $
			parameter has been kept fixed to 3.0 for all the clusters.}
		\begin{tabular}{ccccccc}
			\toprule
			cluster ID & $n_0$ (cm$^{-3}$) & $r_c \ (R_\tup{vir})$ & $\alpha$ & $\beta$ & \
			$r_s \ (R_\tup{vir})$ & $\varepsilon$\\
			\midrule
			46 & $ 0.019 \pm 0.004 $ & $ 0.056 \pm 0.004 $ & $ 0.6 \pm 0.3 $ &
			 $ 0.63 \pm 0.08 $ & $ 0.59 \pm 0.08 $ & $ 2.3 \pm 1.2 $\\
			93 & $ 0.019 \pm 0.005 $ & $ 0.070 \pm 0.005 $ & $ 2.1 \pm 0.2 $ &
			 $ 0.69 \pm 0.07 $ & $ 0.61 \pm 0.07 $ & $ 2.5 \pm 1.3 $\\
			98 & $ 0.018 \pm 0.004 $ & $ 0.064 \pm 0.004 $ & $ 0.5 \pm 0.3 $ &
			 $ 0.7 \pm 0.1 $ & $ 0.6 \pm 0.1 $ & $ 2.6 \pm 1.5 $\\
			103 & $ 0.021 \pm 0.004 $ & $ 0.078 \pm 0.004 $ & $ 1.4 \pm 0.2 $ &
			 $ 0.79 \pm 0.09 $ & $ 0.63 \pm 0.09 $ & $ 2.3 \pm 1.4 $\\
			205 & $ 0.014 \pm 0.002 $ & $ 0.088 \pm 0.002 $ & $ 0.2 \pm 0.2 $ &
			 $ 0.7 \pm 0.1 $ & $ 0.6 \pm 0.1 $ & $ 2.5 \pm 1.4 $\\
			256 & $ 0.020 \pm 0.005 $ & $ 0.082 \pm 0.005 $ & $ 1.9 \pm 0.2 $ &
			 $ 0.79 \pm 0.08 $ & $ 0.64 \pm 0.08 $ & $ 2.6 \pm 1.4 $ \\
			\bottomrule
		\end{tabular}
		\label{tab:nefitparams}
	\end{table*}

	\item \emph{Angular velocity}:
	the complementary work of \citetalias{baldi:angmom} shows that the possible rotation of the gas in
	our cluster sample can be described by a generalized radial profile of the tangential velocity, rather than
	by a solid body model \citepalias[differently from][]{cooray:ksz}.
	We call this law vp2b model following the notation of \citetalias{baldi:angmom},
	and we derive the corresponding radial profile of the angular velocity as:
	\begin{equation}
		\label{eqn:omegavp2b}
		\omega(r) = \frac{\text{vp2b}(r)}{r} = \frac{v_{t0}/r_0}{1 + (r/r_0)^2} \ ,
	\end{equation}
	being $ r_0 $ and $ v_{t0} $ the parameters of the vp2b model, i.e. the scale radius and half of the
	velocity at this radius, $ v_{t0} = \text{vp2b}(r_0)/2 $ \citepalias[see also equation (10) of][]{baldi:angmom}.
\end{itemize}

\begin{figure}
	\centering
	\includegraphics[width=0.4\textwidth]{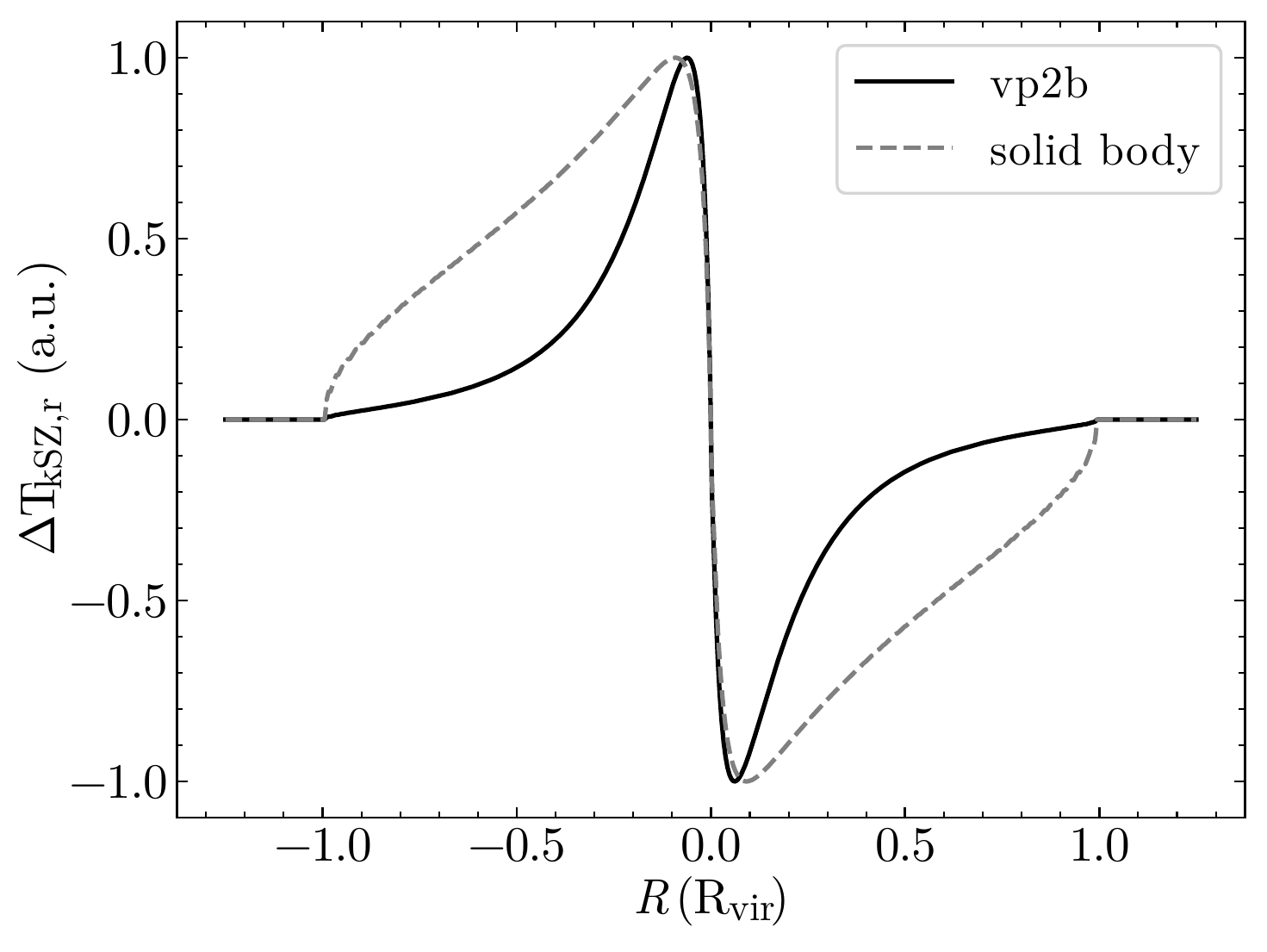}
	\caption{\small Central cuts along the theoretical maps of the rotational kSZ effect, computed using the vp2b and
		a solid body model to derive the angular velocity profile in equation~\eqref{eqn:coorayteo}.
		The two maps have been normalized, in order to highlight the differences in the shape of the cuts along the dipole.}
	\label{fig:comparemodels}
\end{figure}

Fig.~\ref{fig:comparemodels} shows the central cuts along an example theoretical kSZ map
computed according to equation~\eqref{eqn:coorayteo}, using the solid body model -- with constant angular
velocity $ \omega $ -- and the vp2b model, for a fixed profile of the electron number density.
These cuts show that the dipole spots, which have here the same amplitude for comparison purposes, are more broadened
in the case of constant angular velocity. The spatial scale of the dipole, that can be estimated
as the distance between the maximum and the minimum peaks, is of the order of $ \approx 0.2\Rvir $
for both models.
This is because this separation has a stronger dependence on the parameters setting the slopes
of the electron number density profile -- which has been kept fixed here --
rather than on the velocity profile, in agreement with the results from \citet{chluba:ksz}.
%
\subsection{Simulated kSZ maps}
\label{subsec:syntheticmaps}
The kSZ maps of the synthetic clusters in our data set have been produced using the
\texttt{pymsz}\footnote{\texttt{\url{https://github.com/weiguangcui/pymsz}}} package,
which provides mock observations of both the thermal and kinetic SZ effect.
The kSZ temperature signal is computed as:
\begin{equation}
	\label{eqn:simukSZ_cw}
	\frac{\Delta T_\tup{kSZ}(\hat n)}{T_\tup{CMB}} = - \frac{\sigma_T}{c D_A^2} \sum_{i}^{N_P} N_{e,i} \ v_{p,i} \
	W_p(r_i,h_s) \ ,
\end{equation}
where $ D_A $ is the angular diameter distance of the cluster.
The sum extends over the total number of particles along the line of sight, $ N_P $, each being located at a distance
$ r_i $ from the centre of mass, having projected velocity $ v_{p,i} $ and containing $N_{e,i}$ electrons.
The $ W_p $ function is the SPH smoothing kernel of the simulation \citep[see][]{sembolini:music1}.
It is used to smear the kSZ from each gas particle to the projected image pixels, being $ h_s $ the smoothing length of
the gas particles.
We compute the maps as described in equation~\eqref{eqn:simukSZ_cw} in two different versions, to match the only
rotational and the rotational+bulk cases (corresponding to the theoretical prescriptions of
equations~\eqref{eqn:coorayteo} and~\eqref{eqn:coorayteobulk}, respectively). To separate the rotational component,
we simply subtract the cluster bulk velocity, estimated as the average gas particle velocity,
from the velocity of all single gas particles.
In this way we can fit separately the kSZ maps computed from the data to the corresponding model in the two configurations,
in order to establish whether it is possible to recover the expected rotational properties.
Clearly, this is a simplification that cannot be used when dealing with real data, since it is not possible to
separate the bulk component from the total signal, though some complementary methods to estimate the peculiar
velocity could be used \citep[e.g. the Tully-Fisher relation in the case of nearby clusters, see][]{kashlinsky:bulkvelox}.

With the aim to maximise the rotational signal, we choose the best observational configuration to detect the dipole, 
that corresponds to keep the angle between the line of sight and the rotation axis fixed to $ i = 90\dg $
(edge-on with respect to the rotation axis).
To get the corresponding projection, we perform a change of coordinate system, by transforming all the
particle positions and velocities according to a suitable rotation matrix.
Since we assume, as zero-th order approximation, that the rotation axis coincides with the direction of the angular 
momentum vector of the gas computed at the virial radius, we construct the rotation matrix so that the $ z $ axis of
the rotated reference frame coincides with this vector.
With this choice the $ x $ axis of the map should be aligned with the dipole spots, while the $ y $ axis should
correspond to the projection of the rotation axis on the plane of the sky, as also illustrated in Fig.~\ref{fig:mapsketch}.
We want to stress here that we adopt these simplifications just to investigate the detectability of the rotation
in suitable candidate clusters (relaxed with large spin parameter) using the kSZ at the best observational configuration.

In order to validate the effective rotational origin of the dipole pattern, we produce different
projections for each cluster, obtained by selecting different lines of sight, all lying on the orthogonal plane to the
rotation axis, always fulfilling the edge-on condition.
Each line of sight is identified by the angle $ \theta_\tup{los} $, that indicates the separation with respect to the
reference line of sight, having $ \theta_\tup{los} = 0\dg $.
We take a total of six lines of sight, separated by uniform steps of $ \Delta \theta_\tup{los} = 30\dg $,
so that $ 0\dg \leq \theta_\tup{los} \leq 150\dg $.
If a dipole is present because of ICM rotation, its approaching and receding spots should show the same sign and
orientation in the maps, regardless of the particular line of sight chosen for the projection.
As the subtraction of the cluster bulk velocity, this simplification in the interpretation of the results is also
possible only when dealing with data from simulations, since observations can be made along only one line of sight.
%
%

The maps of each cluster extend over $ 2.5\Rvir $ on a side, with a pixel size
$ d_\tup{pix} = \num{5e-3} \Rvir$ that is of the order of $ \approx 10 $ kpc.
For practical reasons, we assume the analysed clusters to be located at $ z=0.05 $, instead of $ z=0 $.
With this choice, according to the cosmological model adopted in the simulation, the angular diameter distance is
200.7 Mpc. The angular size of each pixel is therefore of the order of 10 arcsec, 
and the maps span 2.4 degrees on each side.
To get results that can be useful for possible future applications to data from real experiments,
we reduce the angular resolution of our simulated maps. To this end, we apply a smoothing with a Gaussian
filter having full width at half maximum equal to 20 arcsec, which is compatible with the resolution of currently
operating microwave instruments (e.g. NIKA2 at $ \sim $ \SI{200}{\GHz}).
\begin{figure}
	\centering
	\includegraphics[height=0.84\textheight]{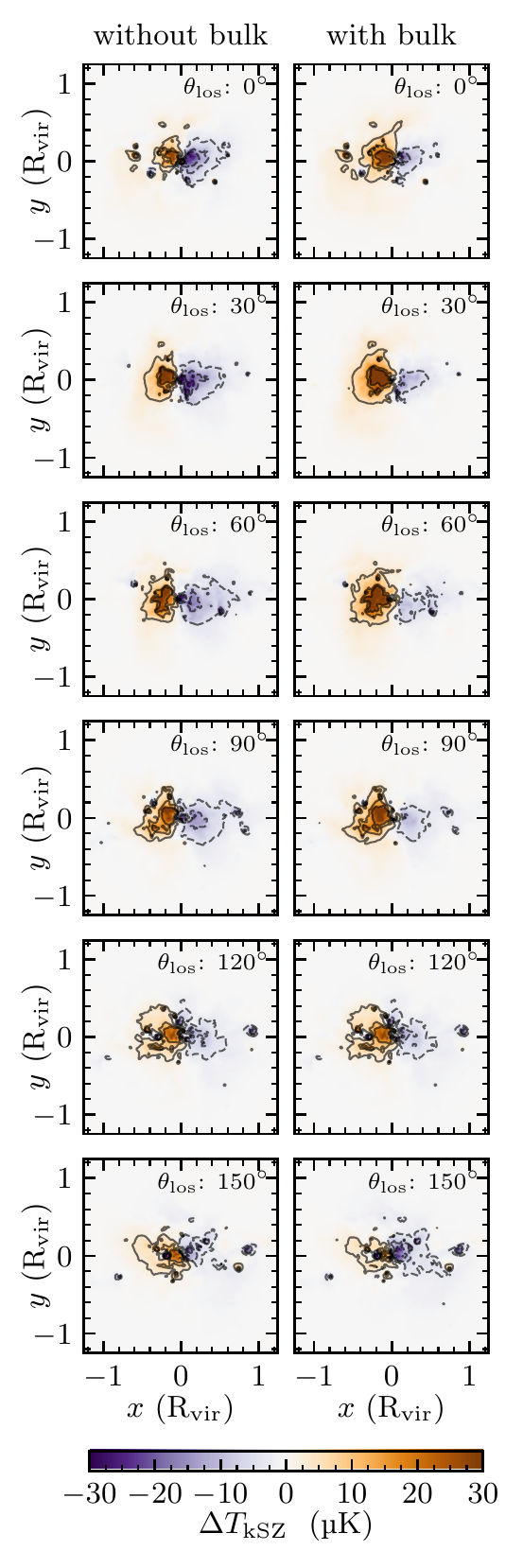} 
	\caption{\small Maps of the temperature shift produced by the kSZ effect for cluster 93, obtained from
		different projections as described in the text, and smoothed at 20 arcsec.
		Left and right panels show the maps obtained without and with the add of the cluster bulk 
		velocity, respectively.
		The angles $ \theta_\tup{los} $ of the corresponding lines of sight are specified on top of each map.
		Contours are plotted from -5$ \sigma $ to 5$ \sigma $, with dashed(solid) lines for negative(positive) values.
		The maximum and minimum values in the maps have been set to $ \pm \SI{30}{\micro\K} $
		for displaying purposes (see colour version of the figure in the online edition).}
	\label{fig:cl93_data}
\end{figure}

Fig.~\ref{fig:cl93_data} shows the kSZ maps for cluster 93 -- which is the most massive cluster in the sample --
smoothed at 20 arcsec, for different lines of sight as described above.
Top panels show the maps generated after the subtraction of the cluster bulk velocity which, on the contrary,
is included in the maps shown in the bottom panels.
It can be seen that, in general, all maps for cluster 93 reported in Fig.~\ref{fig:cl93_data} show a dipole-like
feature with horizontal alignment, and with spots having the same sign at all different projections.
These characteristics confirm that cluster 93 is a good candidate for the inspection of a possible rotation of the ICM
through the kSZ maps. In the maps in the right panels in Fig.~\ref{fig:cl93_data} it is possible
to see how the dipole gets distorted because of the dominating approaching component of the bulk
velocity with respect to the observer for lines of sight having $ \theta_\tup{los} \leq 90\dg $.
The cluster bulk velocity projection is almost null for $ \theta_\tup{los} \geq 120\dg$ instead, so that the
rotational signal remains practically unaffected.
A number of small-scale signal features can be identified in all the maps in Fig.~\ref{fig:cl93_data},
because of the presence of sub-structures.
According to the relaxation criteria that we imposed to select the clusters in our sample (see section~\ref{sec:music}
for details), the masses of these sub-structures are smaller than ten per cent of the mass of the main halo.
They may produce, in some cases, significant outliers in the kSZ temperature maps due to their high
velocity projected on the line of sight.
Despite that, since they extend over scales much smaller than the dipole scale, 
their presence does not affect dramatically the results from the fit to the theoretical maps of the rotational component
of the kSZ. For this reason, the $ \Delta T_\tup{kSZ} $ range in the figures is
set to $ \SI{\pm 30}{\micro\K} $, in order to fit to the dynamic range of the dipole without being affected by
substructure outliers.
The other five clusters in the sample show maps with very similar features, as shown in
Appendix~\ref{sec:appendix}.
\begin{figure}
	\centering
	\includegraphics[width=0.46\textwidth]{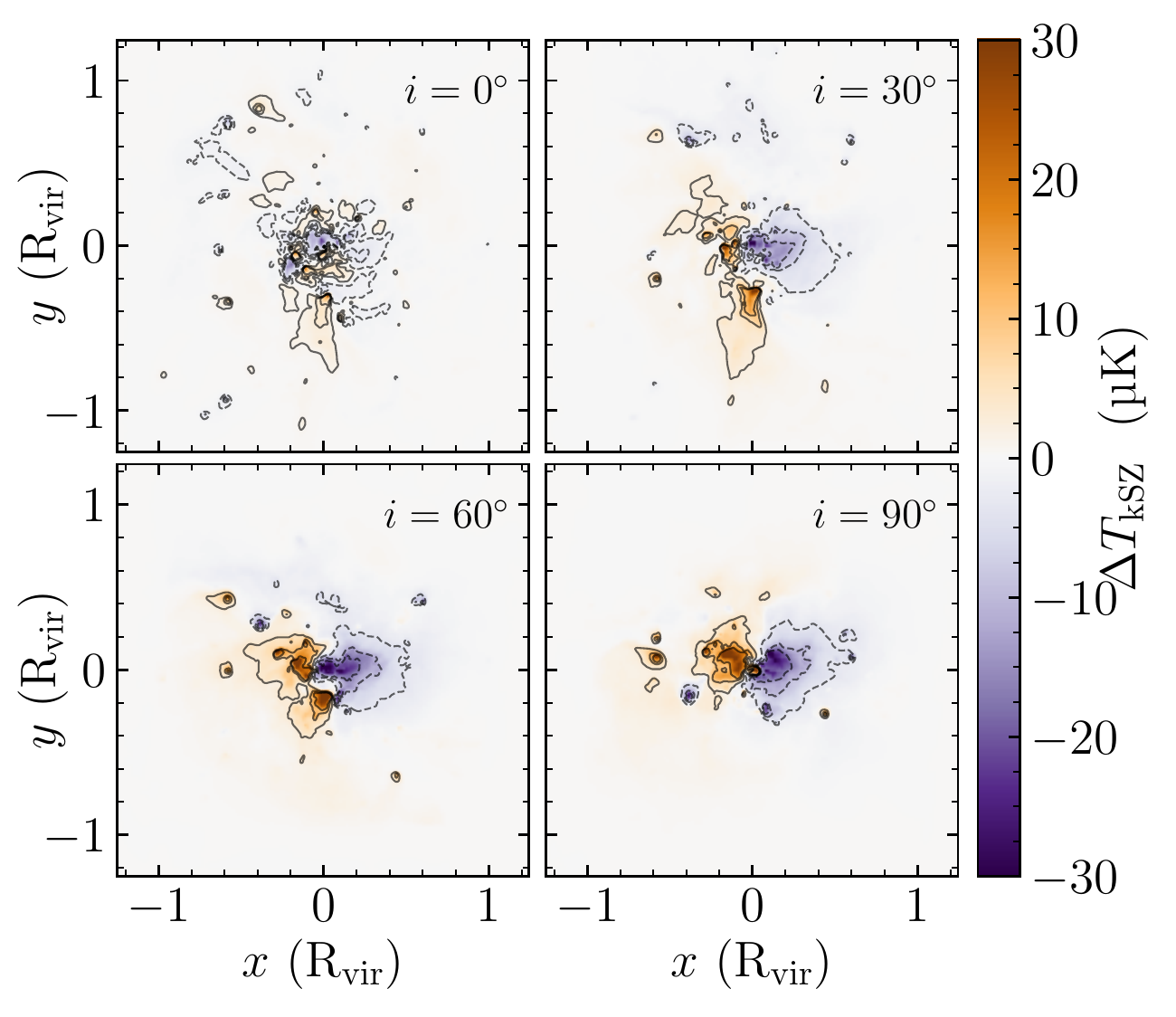}
	\caption{\small Maps of the kSZ effect of cluster 93, for different values of the angle $ i $,
		without accounting for the contribution of the bulk motion.
		The dipole pattern weakens from edge-on to face-on with respect to the rotation axis
		(see colour version of the figure in the online edition).}
	\label{fig:tiltings}
\end{figure}

To highlight the impact of different orientations of the line of sight with respect to the rotation axis,
we show in Fig.~\ref{fig:tiltings} the maps of cluster 93 at $ \theta_\tup{los} = 0\dg $ taken at different angles $ i $.
We verify that the dipole is clearly visible in the case of orthogonal line of sight with respect to the rotation axis,
with decreasing amplitude for decreasing values of $ i $, consistently with the expectations.
Unfortunately, the contribution from the $ \sin i $ term and the $ v_{t0} $ parameter cannot be discriminated
in the observed signal. For this reason we set $ i=90\dg $ in the rest of the analysis.

%% file: results.tex
\section{Results and discussion}
\label{sec:results}
In order to recover the rotational properties of our test clusters, we use the Levenberg-Marquardt least-square
algorithm \citep{more:lmalgorithm} to compute a pixel-to-pixel fit to the synthetic kSZ maps described
in section~\ref{subsec:syntheticmaps}.
We treat separately the purely rotational case, referring to the theoretical model given by equation~\eqref{eqn:coorayteo},
and the full case accounting also for the cluster bulk motion, referring to the model of equation~\eqref{eqn:coorayteobulk}.
The six parameters of the radial profile of the electron number density are kept fixed, with the values
listed in Table~\ref{tab:nefitparams} of section~\ref{subsec:theoreticalmaps}.
The free parameters we recover from the fit to the kSZ maps are the scale radius, $ r_0 $, and the
scale velocity, $ v_{t0} $, introduced in equation~\eqref{eqn:omegavp2b}.
In the full case including the bulk motion, we treat the cluster bulk velocity projected on the line of sight
($ v_\tup{bulk} $) as an additional free parameter.
In order to account for a possible non-horizontal alignment of the dipole with respect to the 
projection of the rotation axis ($ y $ axis in the maps), we repeat the fit also adding another free 
parameter to the theoretical
model of the kSZ map.
It consists in an offset, $ \delta \phi $, added to the azimuthal angle $ \phi $, which partially 
accounts
for the error in the estimate of the correct orientation of the rotation axis of the ICM.
If $ \delta \phi = 0\dg $, it is correct to assume the direction of the angular momentum of the
gas at virial radius as the rotation axis. We verify that this offset takes values smaller than $ 10\dg $ in most cases,
and that it does not affect significantly the results from the fit.
For this reason, we refer only to the case with $ \delta \phi = 0\dg $ in all the forthcoming results.
\begin{figure*}
	\centering
	\subfloat[without bulk motion]{\includegraphics[height=0.90\textheight]{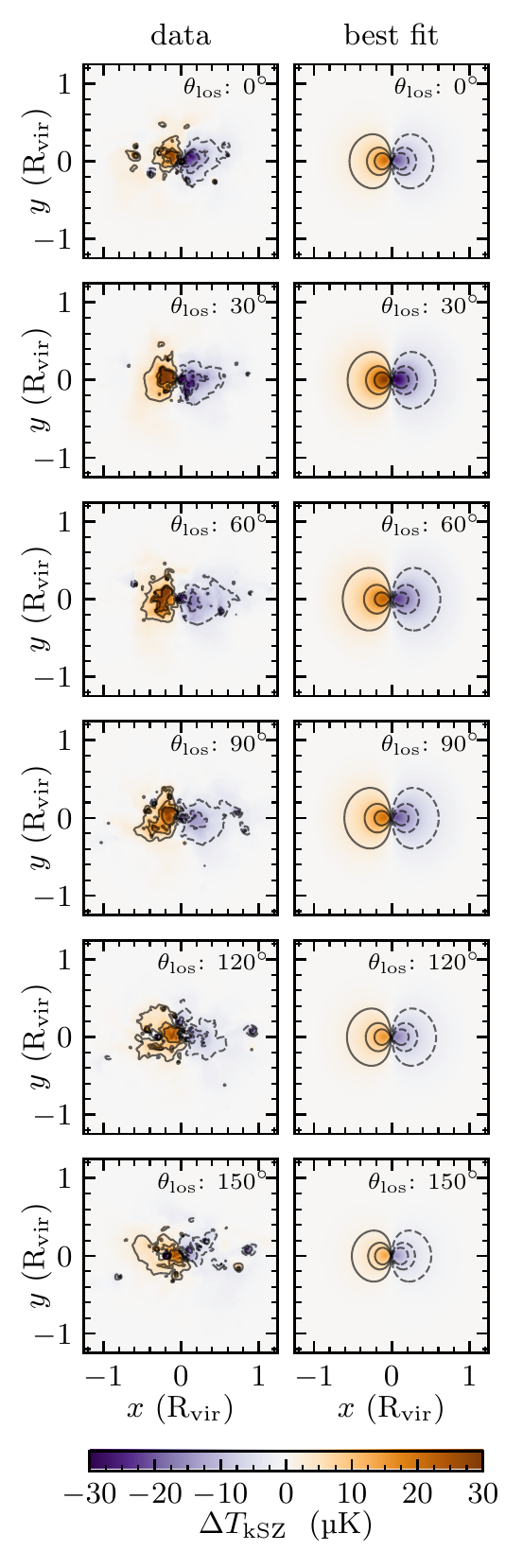} 
		\label{fig:cl93_datafits_nobulk}}\qquad
	\subfloat[with bulk motion]{\includegraphics[height=0.90\textheight]{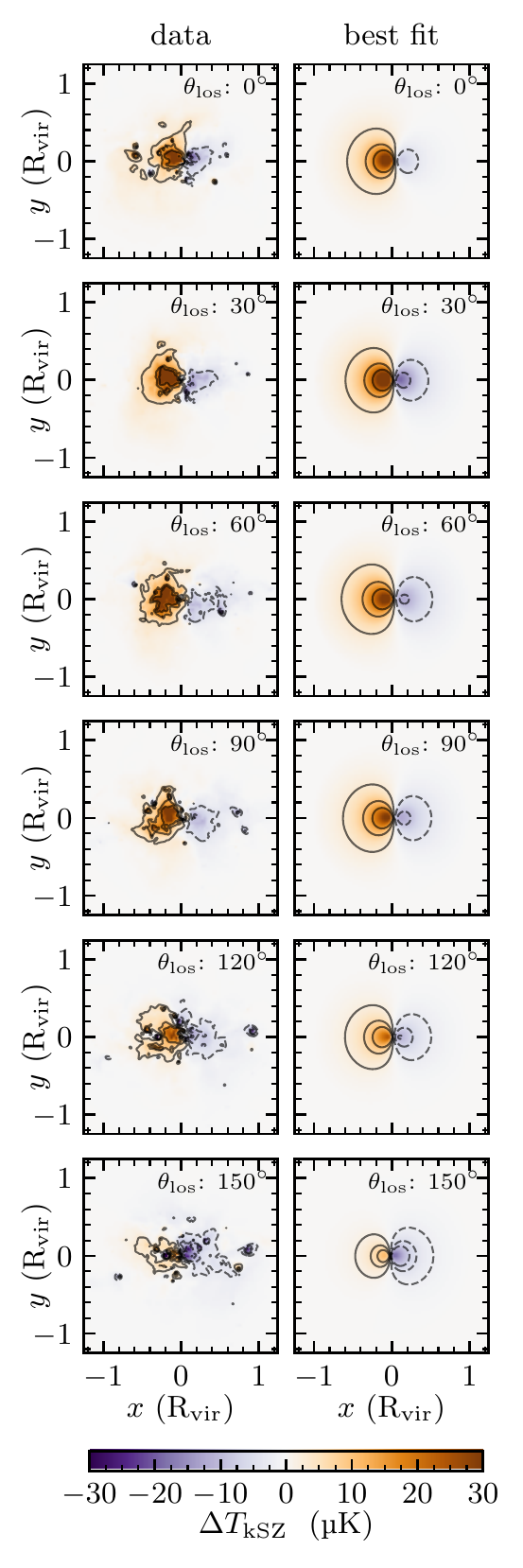} %
		\label{fig:cl93_datafits_bulk}}
	\caption{\small Comparison between the kSZ temperature maps of cluster 93 and the corresponding
		best-fit maps, without and with accounting for the cluster bulk velocity (left and right
		panels, respectively).
		Contours are plotted from -5$ \sigma $ to 5$ \sigma $, with dashed(solid) lines for negative(positive)
		values. The colorbar is set to $ \pm \SI{30}{\micro\K} $ for displaying purposes
		(see colour version of the figure in the online edition).}
	\label{fig:cl93_datafits}
\end{figure*}
\begin{figure}
	\centering
	\subfloat[without bulk motion]{\includegraphics[width=0.48\textwidth]{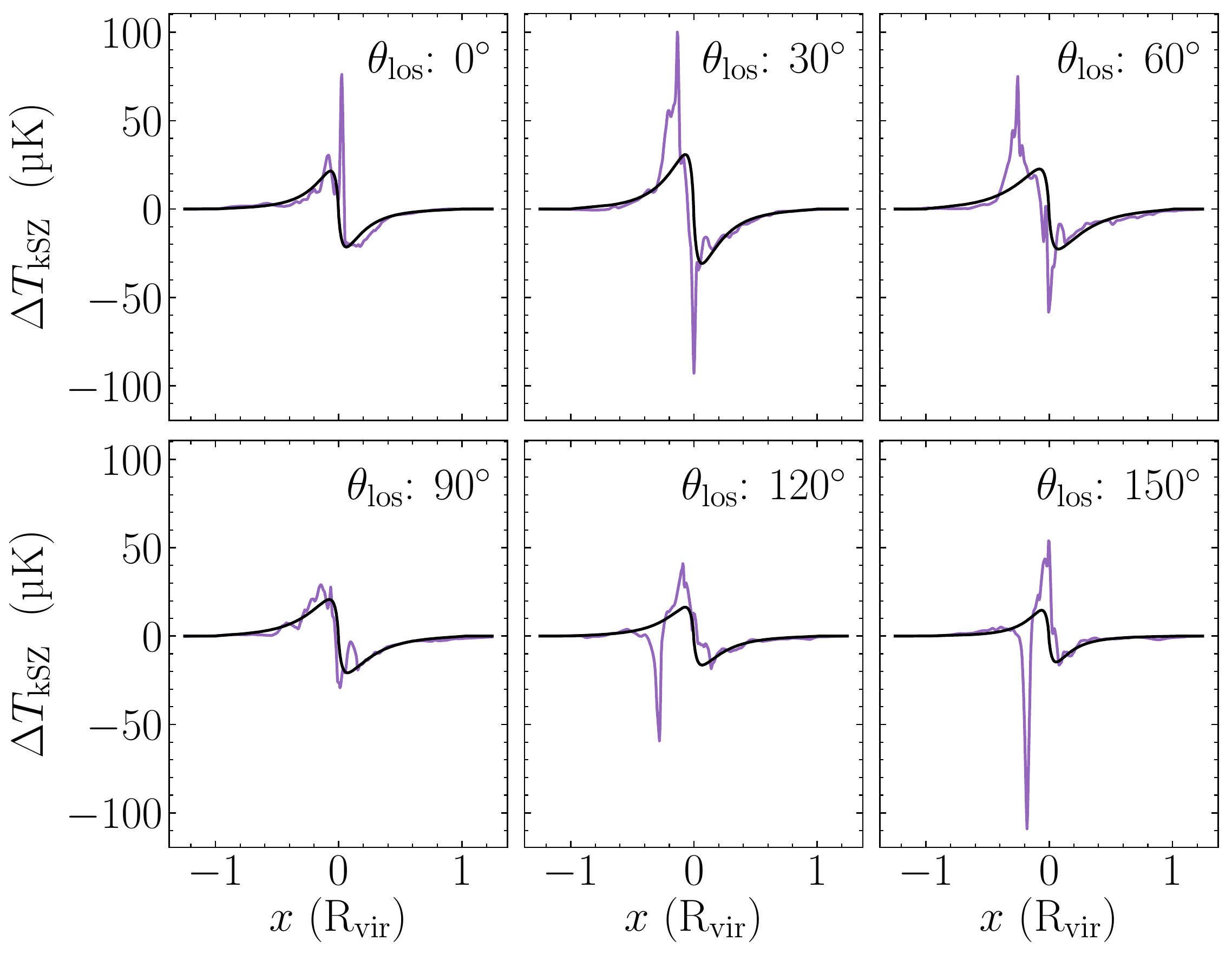} %
		\label{fig:cl93_cuts_nobulk}}\quad
	\subfloat[with bulk motion]{\includegraphics[width=0.48\textwidth]{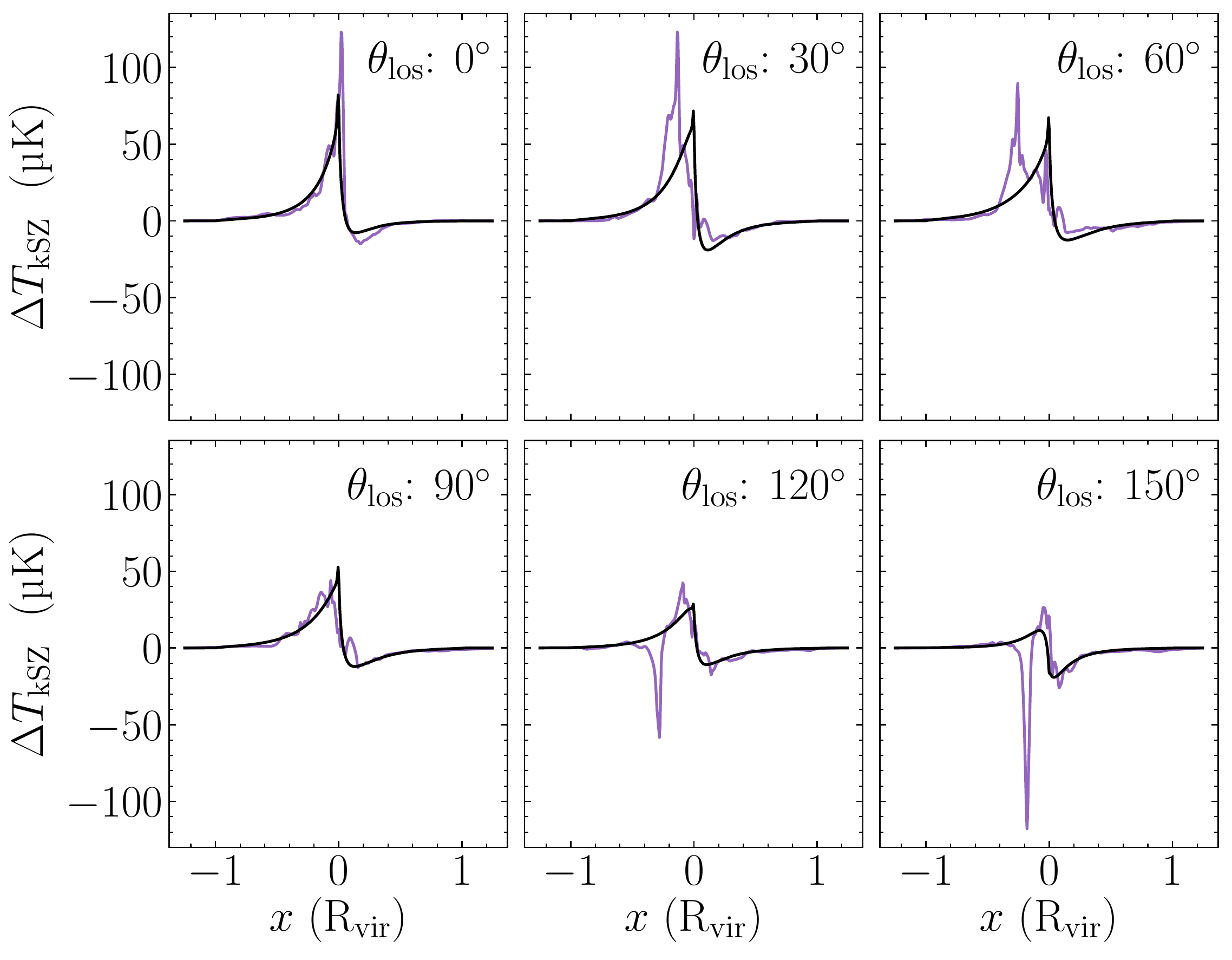} %
		\label{fig:cl93_cuts_bulk}}
	\caption{\small Central cuts from the kSZ maps of cluster 93, without (top panel) and with
		(bottom panel)
		the cluster bulk velocity, for different lines of sight.
		Purple curves represent the cut through the data maps, while black curves represent the
		cut through the best-fit maps. The presence of high-velocity particles can be seen as
		outliers at small scales.}
	\label{fig:cl93_cuts}
\end{figure}

Fig.~\ref{fig:cl93_datafits} shows the kSZ maps generated from the data with the corresponding best-fit
theoretical maps, all smoothed at 20 arcsec, without (left panel) and with (right panel)
the bulk motion, in the case of cluster 93.
Similar maps for the complete sample showing the best lines of sight are reported in Appendix~\ref{sec:appendix}.
We also show in Fig.~\ref{fig:cl93_cuts} -- for cluster 93 as before -- the central cuts through the same
maps of Fig.~\ref{fig:cl93_datafits}, to better highlight how our procedure recovers the features of the signal.
It can be seen that, in general, the theoretical model is appropriate to describe the data in both cases of
subtraction and adding of the cluster bulk velocity.
The signal in the data maps may be larger of a factor of $ \approx 3$ at most with respect to the best-fit maps
in some regions, because of small-scale outliers due to sub-structures (that can be clearly spotted in the
plots of Fig.~\ref{fig:cl93_cuts}), whose contribution generally changes from one line of sight to another (see also
Fig.~\ref{fig:cl93_data}).
\begin{table*}
	\centering
	\caption{\small Best-fit values of the free parameters $ r_0 $ (scale radius) and $ v_{t0} $ (scale velocity)
		of equation~\eqref{eqn:omegavp2b}. Left and middle columns report the average values with their standard deviation
		computed over all the lines of sight, as derived from the fit to the kSZ maps without and with the bulk term, 
		respectively. Right columns lists the expected values for the parameters, derived from the fit to the
		tangential velocity data, $ v_t $, extracted from the simulation. Velocities are given in units of the
		circular velocity at the virial radius, $ v_\tup{circ} = \sqrt{G \Mvir/\Rvir} $.}
	\begin{tabular}{ccccccc}
		\toprule
		\multirow{2}{*}{cluster ID} &
		\multicolumn{2}{c}{fit to kSZ without bulk} &
		\multicolumn{2}{c}{fit to kSZ with bulk} &
		\multicolumn{2}{c}{fit to $ v_t $ data}\\
		\cmidrule{2-7}
		& $ r_0 \ (\Rvir)$ & $ v_{t0} \ (v_\tup{circ}) $ &
		$ r_0 \ (\Rvir)$ & $ v_{t0} \ (v_\tup{circ}) $ &
		$ r_0 \ (\Rvir)$ & $ v_{t0} \ (v_\tup{circ}) $\\
		\midrule
		46 & $0.34 \pm 0.09$ & $0.84 \pm 0.04$ & 
		$0.37 \pm 0.05$ & $0.95 \pm 0.17$ &
		$ 0.36 \pm 0.22 $ & $ 0.56 \pm 0.17 $\\
		93 & $0.34 \pm 0.08$ & $0.87 \pm 0.27$ &
		$0.34 \pm 0.07$ & $0.88 \pm 0.28$ &
		$ 0.49 \pm 0.27 $ & $ 0.58 \pm 0.20 $\\
		98 & $0.38 \pm 0.30$ & $1.53 \pm 0.67$ &
		$0.33 \pm 0.25$ & $1.39 \pm 0.65$ &
		$ 0.57 \pm 0.62 $ & $ 0.49 \pm 0.29 $\\
		103 & $0.29 \pm 0.13$ & $0.82 \pm 0.09$&
		$0.28 \pm 0.11$ & $0.83 \pm 0.09$ &
		$ 0.47 \pm 0.35 $ & $ 0.52 \pm 0.19 $\\
		205 & $0.20 \pm 0.08$ & $0.96 \pm 0.08$ &
		$0.24 \pm 0.09$ & $0.86 \pm 0.19$ &
		$ 0.27 \pm 0.14 $ & $ 0.65 \pm 0.17 $\\
		256 & $0.32 \pm 0.12$ & $1.00 \pm 0.23$ &
		$0.33 \pm 0.11$ & $1.02 \pm 0.24$ &
		$ 0.37 \pm 0.20 $ & $ 0.61 \pm 0.22 $\\
		\bottomrule
	\end{tabular}
	\label{tab:results_all}
\end{table*}
Table~\ref{tab:results_all} lists the $ r_0 $ and $ v_{t0} $ parameters estimated from the fit.
Since they should have consistent values independently on the observed projection, we report the average and
standard deviation over the different lines of sight we considered.
We find that the values of both parameters are in agreement within one standard
deviation when comparing the two cases of subtraction and add of the bulk term
(listed in the left and centre columns of Table~\ref{tab:results_all}, respectively).
This indicates that, in principle, this procedure is able to disentangle the signal produced by rotation from the
one given by the bulk motion.
The comparison between these results and the values derived from the fit to the velocity inferred directly from the
simulation data (listed in the right columns in Table~\ref{tab:results_all}), shows that the
scale radius $ r_0 $ is consistent within one standard deviation.
The values of the $ v_{t0} $ parameter are overestimated with respect to the expected ones by a factor
of $ \sim 1.5(1.6) $ on average, when subtracting(adding) the bulk motion. Indeed, they reach fractions larger
than 80 per cent of the circular velocity at the virial radius (which is $ \gtrsim \SI{1000}{\km\per\second} $ for
all the clusters). Nevertheless, there is agreement within one standard deviation for almost all clusters.
A noticeable exception is given by cluster 98, for which we get a larger overestimate (of factors 3.1 and 2.8
without and with the cluster bulk, respectively), and significantly larger errors.
Such discrepancies may be due, in general, to a less efficient reconstruction of the dipole because
of a higher impact from irregularities in the gas density distribution, and from high-velocity sub-structures,
especially in the outer regions (see e.g. the corresponding panels in Fig.~\ref{fig:data_allos_allclus}).
\begin{table}
	\centering
	\caption{\small Amplitude of the kSZ temperature signal measured from the best-fit maps.
		The $ A_\tup{dip} $ column refers to the amplitude of the dipole averaged over all the lines of sight,
		as derived from the fit to the maps without the bulk motion. $ A_\tup{bulk} $ refers
		instead to the maximum amplitude of the best fit to the maps accounting for the cluster bulk motion.}
	\begin{tabular}{ccc}
		\toprule
		cluster ID & $ A_\tup{dip} $ (\si{\micro\K}) & $ A_\tup{bulk} (\si{\micro\K}) $\\
		\midrule
		 46 & $ 10.8 \pm 2.5 $ &  -57.5 \\ 
		 93 & $ 21.1 \pm 5.2 $ & 82.1  \\ 
		 98 & $ 24.4 \pm 9.2$ &  -77.9 \\ 
		103 & $ 16.5 \pm 2.9 $ &  -99.4  \\ 
		205 & $ 20.9 \pm 5.1 $ &  68.8  \\ 
		256 & $ 24.4 \pm 4.3 $ &  -143.1\\ 
		\bottomrule
	\end{tabular}
	\label{tab:amplitudes}
\end{table}
In order to give a quantitative indication of the kSZ signal coming from rotation, we measure the
amplitude of the dipole from the best-fit theoretical maps in the rotation-only case.
Values are listed in Table~\ref{tab:amplitudes}, as derived from the average over all the explored projections.
It can be seen that the dispersion across different lines of sight is of the order of 38 per cent at most,
while average values are of the order of few tens of \si{\micro\K}. This result is in agreement with the
predictions by \citet{chluba:ksz}, that indicates dipole amplitudes ranging between
$ \sim 4 $ and $ \sim \SI{30}{\micro\K} $ (assuming a solid body rotation).
Table~\ref{tab:amplitudes} also reports the maximum amplitude, $ A_\tup{bulk} $,
measured in the best-fit maps accounting for the cluster bulk motion. We verify that, as expected, this
quantity is linearly proportional to the projected bulk velocity, $ v_\tup{bulk} $.
Using these values it is possible to compute the ratio $ A_\tup{dip}/A_\tup{bulk} $, in order to estimate
how much does the rotation contribute to the total kSZ signal in the case in which the projection of the cluster
bulk velocity takes its maximum value.
The average value of this ratio is $ \sim 0.23 $, confirming that very high sensitivities are needed to measure
the effect of a rotation, provided the best observational conditions.
\begin{figure}
	\centering
	\includegraphics[width=0.48\textwidth]{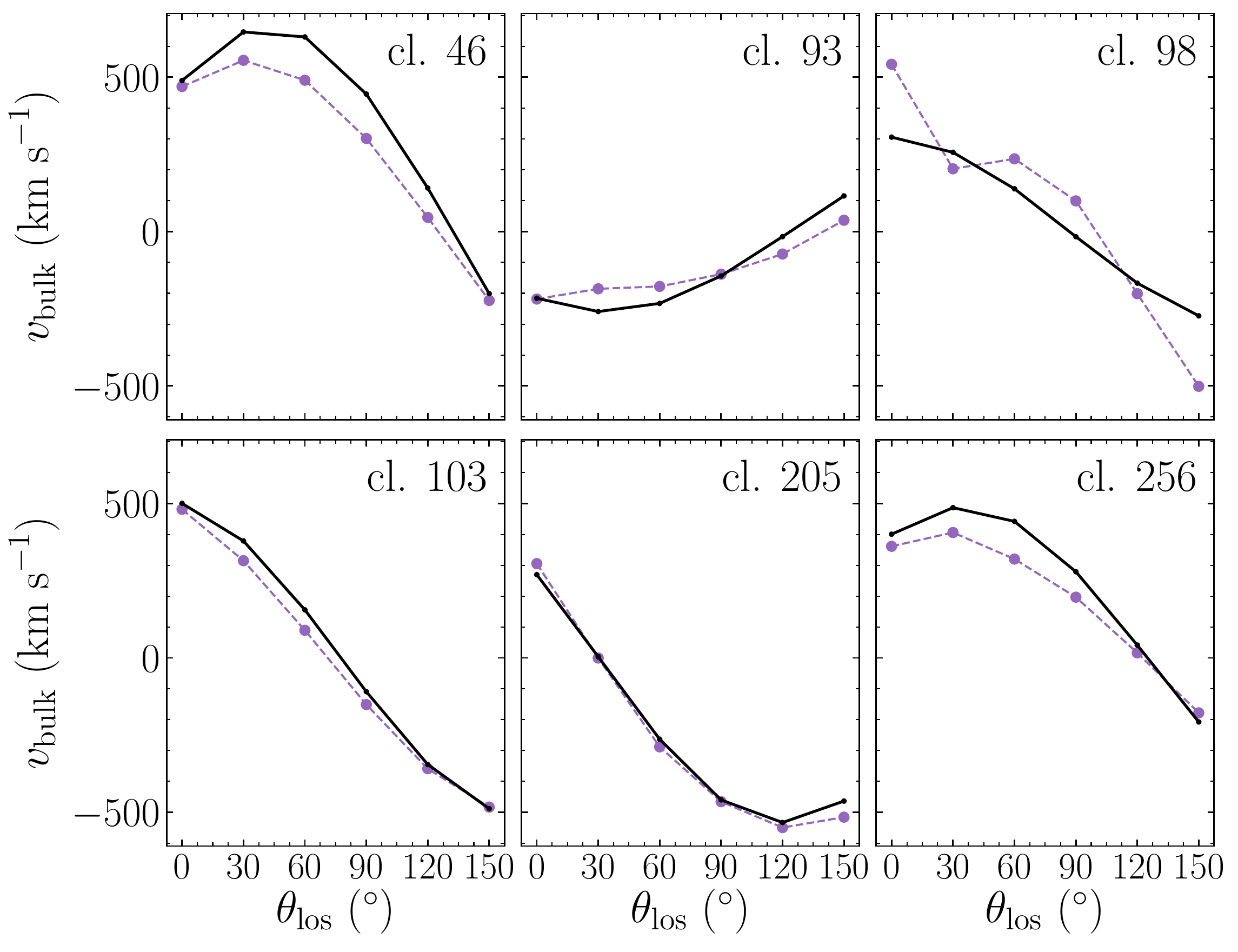} %
	\caption{\small Values of the projected cluster bulk velocity, $ v_\tup{bulk} $, as a function of the
		angle $ \theta_\tup{los} $ identifying the different lines of sight. Purple dots with dashed line
		represent the estimate from the fit to the maps with the bulk term. Black dots with solid line are
		the true values extracted from the simulation.}
	\label{fig:vbulk_vs_los}
\end{figure}
\begin{table}
	\centering
	\caption{\small Normalized difference between the recovered $ v_\tup{bulk} $ and the true one
		estimated from the simulation, $ v_\tup{bulk,sim} $. Values along the columns refer to the six
		different lines of sight.}
	\begin{tabular}{ccccccc}
		\toprule
		\multirow{3}{*}{cluster ID}
		& \multicolumn{6}{c}{$ v_\tup{bulk}/v_\tup{bulk,sim} - 1 $}\\
		\cmidrule{2-7}
		 & \multicolumn{6}{c}{$ \theta_\tup{los} $}\\
		 & $ 0\dg $ & $ 30\dg $ & $ 60\dg $ & $ 90\dg $ & $ 120\dg $ & $ 150\dg $\\
		\midrule
		46  & -0.04  & -0.14  & -0.22  & -0.32  & -0.68  & -0.11\\
		93  & -0.01 &  0.28 &  0.24 &  0.04  & -3.39  & -0.68\\
		98  & 0.77  & -0.21 &  0.70  & -4.96 &  -0.20  & -0.84\\
		103 &  -0.04  & -0.17  & -0.43  & -0.38  & -0.04  & 0.01\\
		205 & 0.13 &  -0.92  & -0.09  & -0.01  & -0.03  & -0.11\\
		256 &  -0.10  & -0.17  & -0.28  & -0.30 & -0.61 & 0.14\\
		\bottomrule
	\end{tabular}
	\label{tab:vbulkz}
\end{table}
The recovered values of $ v_\tup{bulk} $ are of the order of hundreds of \si{\km\per\second}, and
they are fairly compatible with the true values from the simulation.
We show their comparison in Fig.~\ref{fig:vbulk_vs_los}, where it can also be seen the sinusoidal behaviour of
the different projections with varying $ \theta_\tup{los} $.
In Table~\ref{tab:vbulkz} we report the differences between $ v_\tup{bulk} $ recovered from the fit to the kSZ
maps and the true value from the simulation, $ v_\tup{bulk,sim} $, normalized to $ v_\tup{bulk,sim} $ itself.
Values are generally underestimated by few tens of per cent at most projections; differences are more significant
for lines of sight in which the projected bulk velocity takes absolute values smaller than \SI{100}{\km\per\second}.
The possibility of recovering the bulk velocity term with this procedure is an important result of this work.
Indeed, the use of the rotational kSZ effect with complementary observables, e.g. higher order corrections
terms to the Kompaneets approximation, or the degree of CMB polarization induced by the kSZ, could give
an estimate of the three-dimensional cluster velocity \citep{birkinshaw:articolo}.

The simplifying assumptions we have made, e.g. the orthogonal orientation of the los with respect to the rotation axis,
and the poor error constraints we get in the final estimates of the parameters could be
limiting factors for this analysis. Nevertheless, this study is intended to quantify the amount of kSZ signal
that would be produced by ICM rotation at the best observational conditions, also to get a possible validation of the
vp2b model. Some enhancements to get more robust results can include e.g. a proper modelling of the sub-structures
to be included in the theoretical map for a more accurate fit.
Also, a deeper approach focused on a more quantitative assessment of the feasibility of such challenging
observations at millimetre wavelengths is ongoing, by accounting for the full instrumental effects and
contamination from astrophysical sources and SZ background.

%% file: conclusions.tex
\section{Summary and conclusions}
\label{sec:conclusions}
In this paper we address the study of the rotation of the ICM in simulated galaxy clusters through the maps of the
kSZ effect.
In this preliminary analysis, we select a sample of six particularly relaxed and possibly rotating objects from
MUSIC simulations studied in \citetalias{baldi:angmom}, and we fit their kSZ temperature maps to a theoretical model
based on the one proposed by \citetalias{cooray:ksz}.
We adopt a simplified Vikhlinin model for the electron number density, and the vp2b model of
\citetalias{baldi:angmom} for the angular velocity.
We study both a simplified case which does not account for the cluster bulk motion, in order to get only the rotational 
signal, and the complete case that includes an additional term depending on the cluster bulk velocity projected on the
line of sight. We explore six different lines of sight for both cases, picking them all orthogonal to the rotation axis,
in order to maximise the amplitude of rotational signal.
The main results from the fit to our maps can be summarized as follows:
\begin{itemize}
	\item with our procedure we can recover the parameters of the radial profile of the rotational velocity
	within one standard deviation in the case of the scale radius, and within two standard deviations at most
	in the case of the scale velocity, by averaging over all the lines of sight.
	These results are poorly affected by the small-scale outliers produced by high-velocity sub-structures
	located within the cluster virial radius, at most projections;
	\item the amplitude of the best-fit dipole is consistent with the estimates found in the literature for
	relaxed systems. From a comparison between the amplitude derived without and with the bulk motion, we estimate
	that the rotational contribution to the total kSZ signal is, on average, of the order of 23 per cent
	in the best observational conditions;
	\item the projection of the bulk velocity on the line of sight estimated from the full kSZ model shows
	differences of few per cent at most projections with respect to the true values from the simulation.
	As expected, it has a sinusoidal behaviour as a function of the angle identifying each particular line of sight.
\end{itemize}
We plan to refine the analysis presented in this work by accounting for the full pipeline of NIKA2,
including instrumental and astrophysical contaminants, to possibly apply this study to observations of real clusters.
This work is part of a larger project on the feasibility of the characterization of ICM rotation, that will include
the complementary analysis of multi-frequency data generated from MUSIC clusters.
In particular, we are working on the analysis of the dynamics of galaxy members in our cluster sample, to investigate
the possible co-rotation between baryonic components.
In this context, the future \textsl{Euclid} mission \citep{euclid:presentation} could open new interesting possibilities
for observations in the optical band.
We also plan to account for X-ray data, given their complementarity to the SZ effect to derive cluster properties
and dynamics.
For instance, the upcoming X-ray satellite \textsl{Athena} \citep{athena:presentation} will feature an unprecedented
sensitivity and resolution for the characterization of spectral lines, that would turn to be extremely useful to
constrain the ICM kinetic properties.
Lastly, given the recent developments of the \texttt{pymsz} code, another possible future development of the present
work applied to synthetic clusters would include the treatment of the full signal from tSZ and kSZ, to recover e.g.
cluster properties and their impact on cosmology.

%% file: appendix.tex
\section{Maps for the whole sample}
\begin{figure*}
	\centering
	\subfloat[cl. 46]{\includegraphics[width=0.50\textwidth]{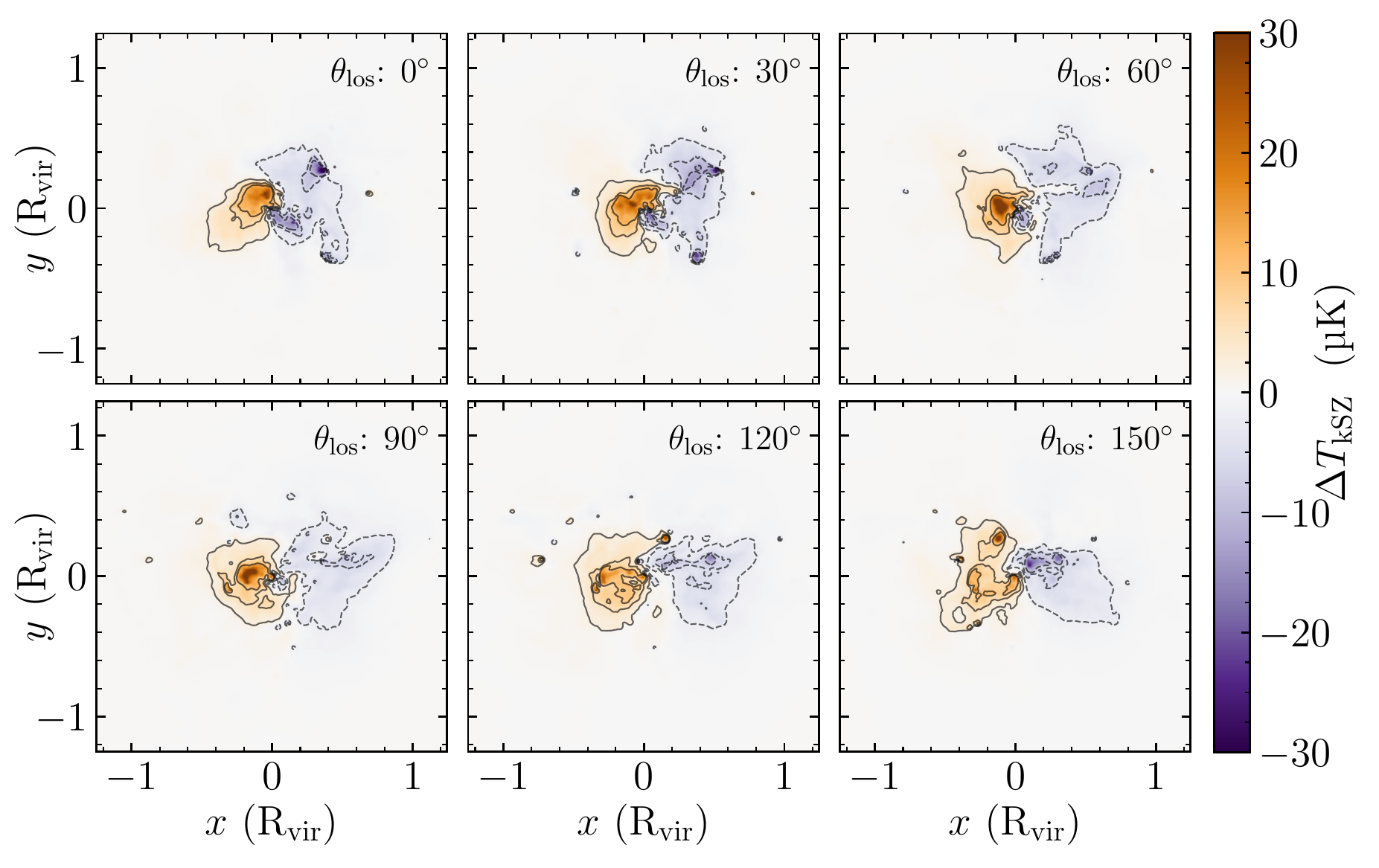}}
	\subfloat[cl. 93]{\includegraphics[width=0.50\textwidth]{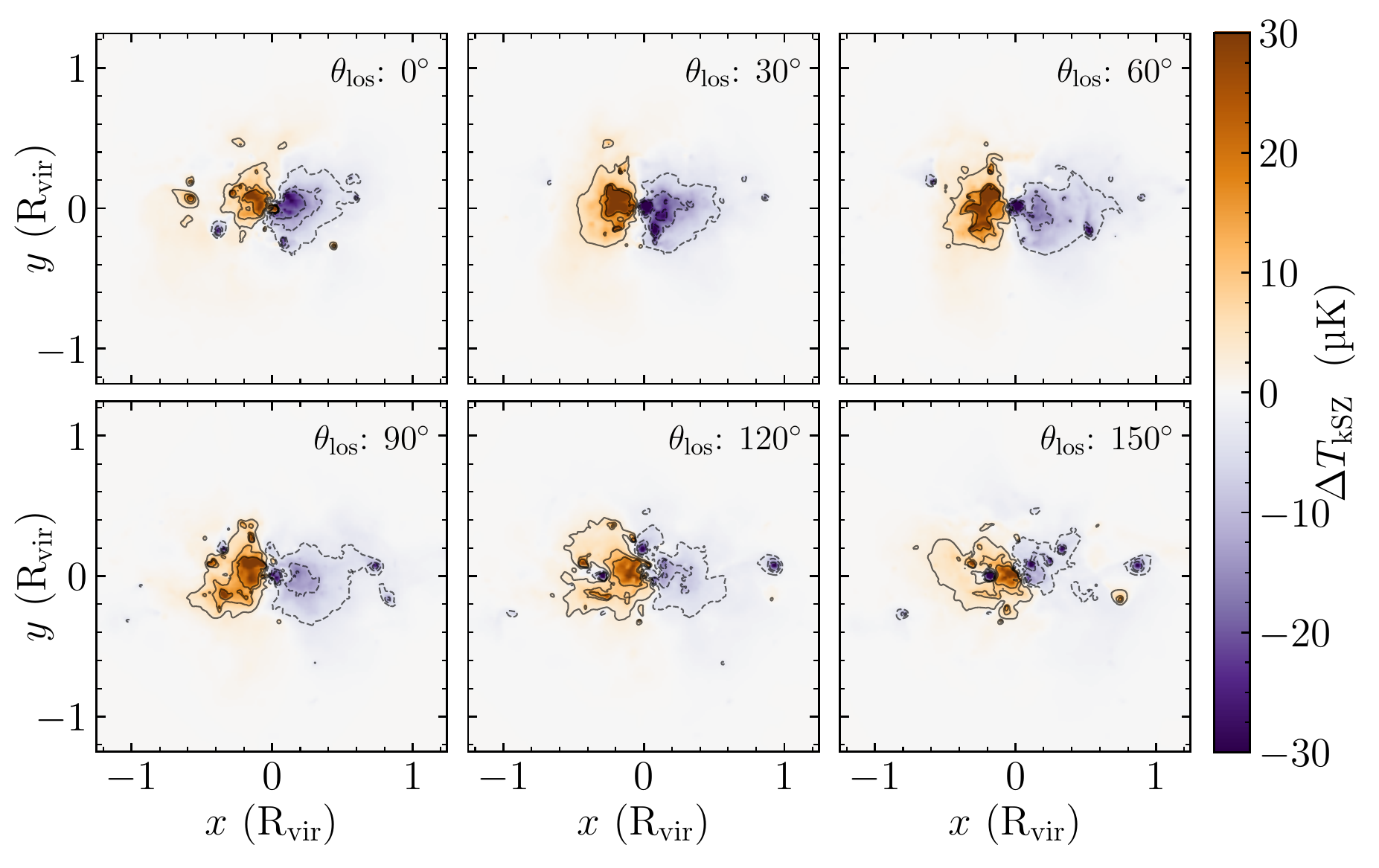}}\qquad
	\subfloat[cl. 98]{\includegraphics[width=0.50\textwidth]{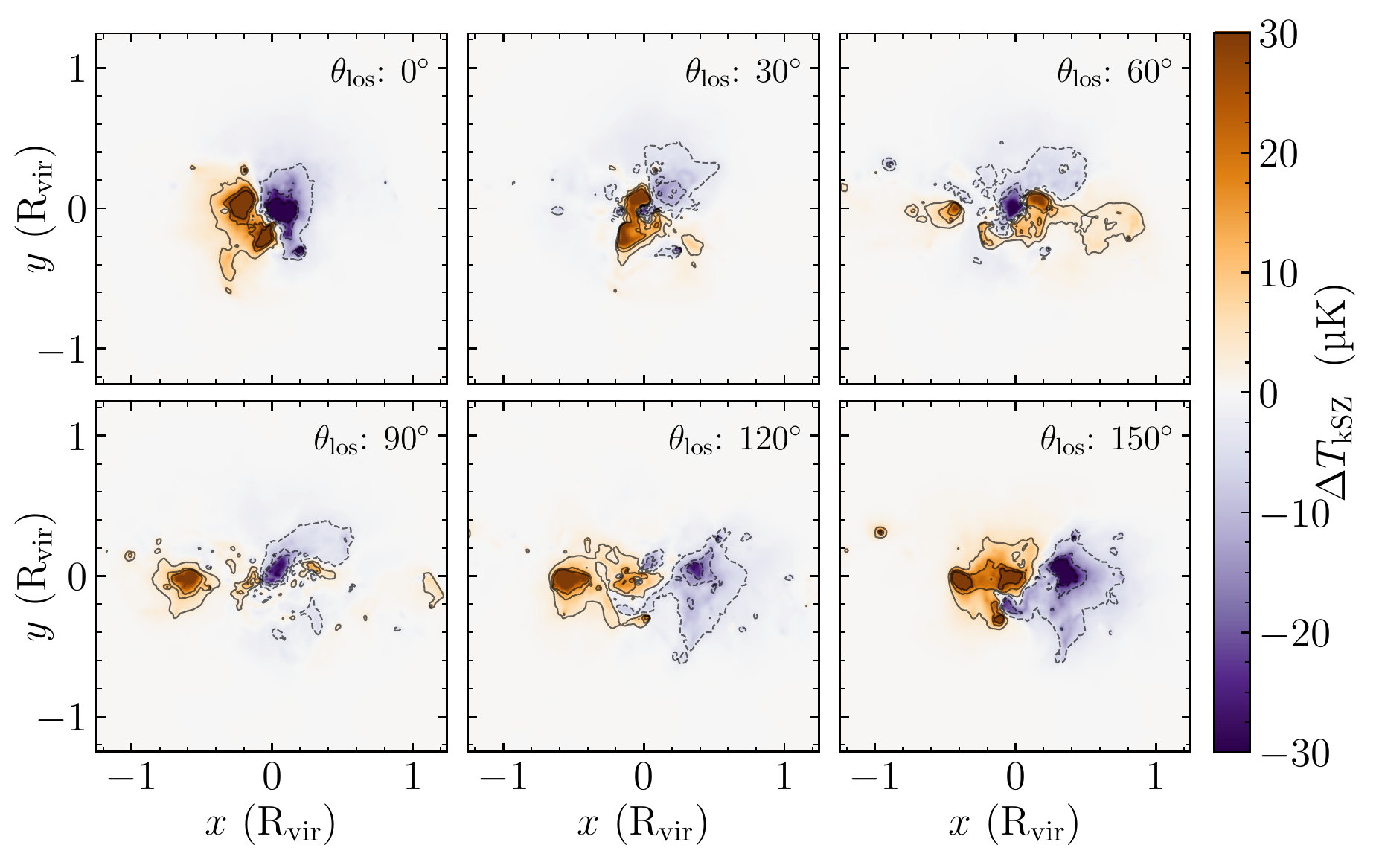}}
	\subfloat[cl. 103]{\includegraphics[width=0.50\textwidth]{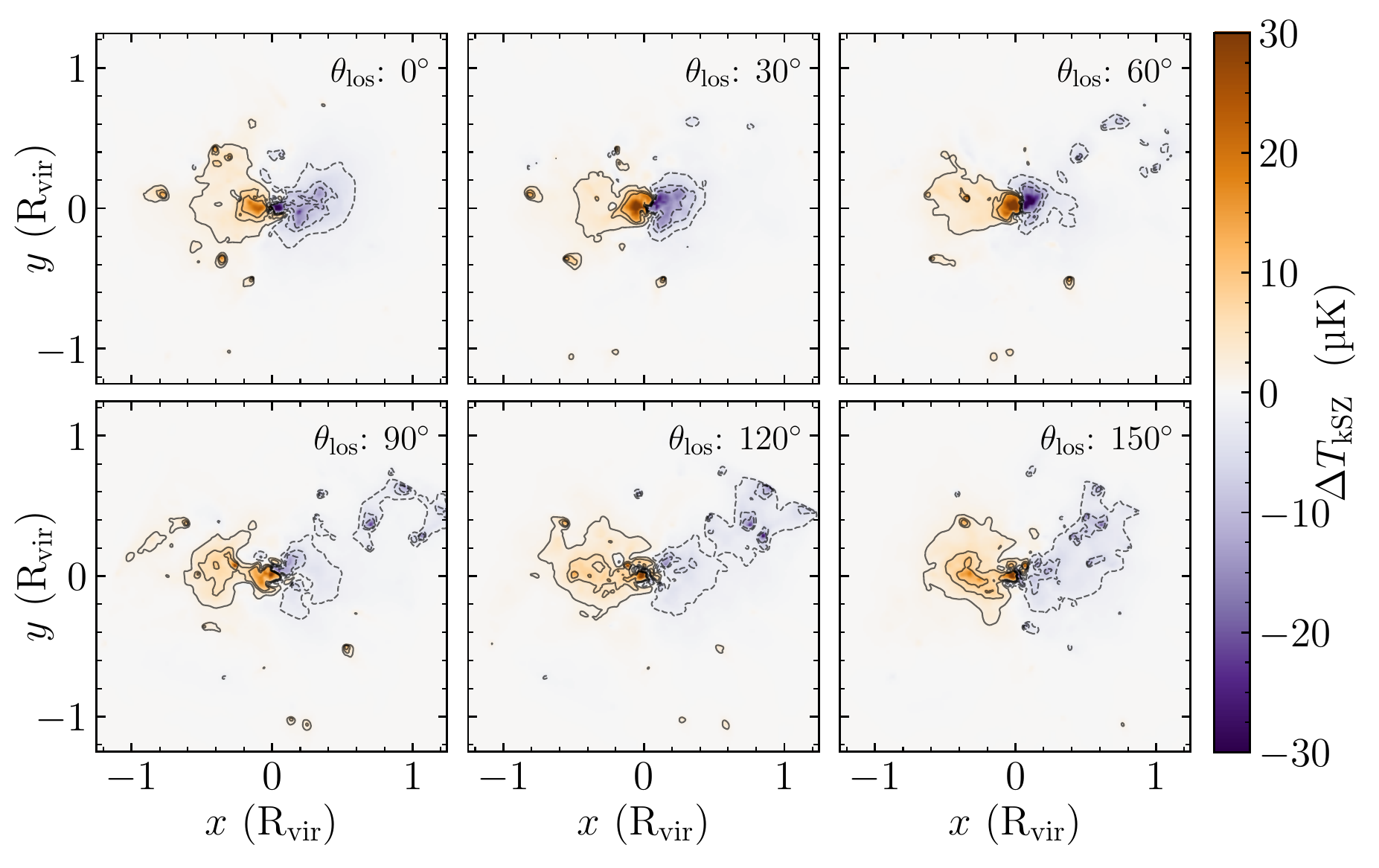}}\qquad
	\subfloat[cl. 205]{\includegraphics[width=0.50\textwidth]{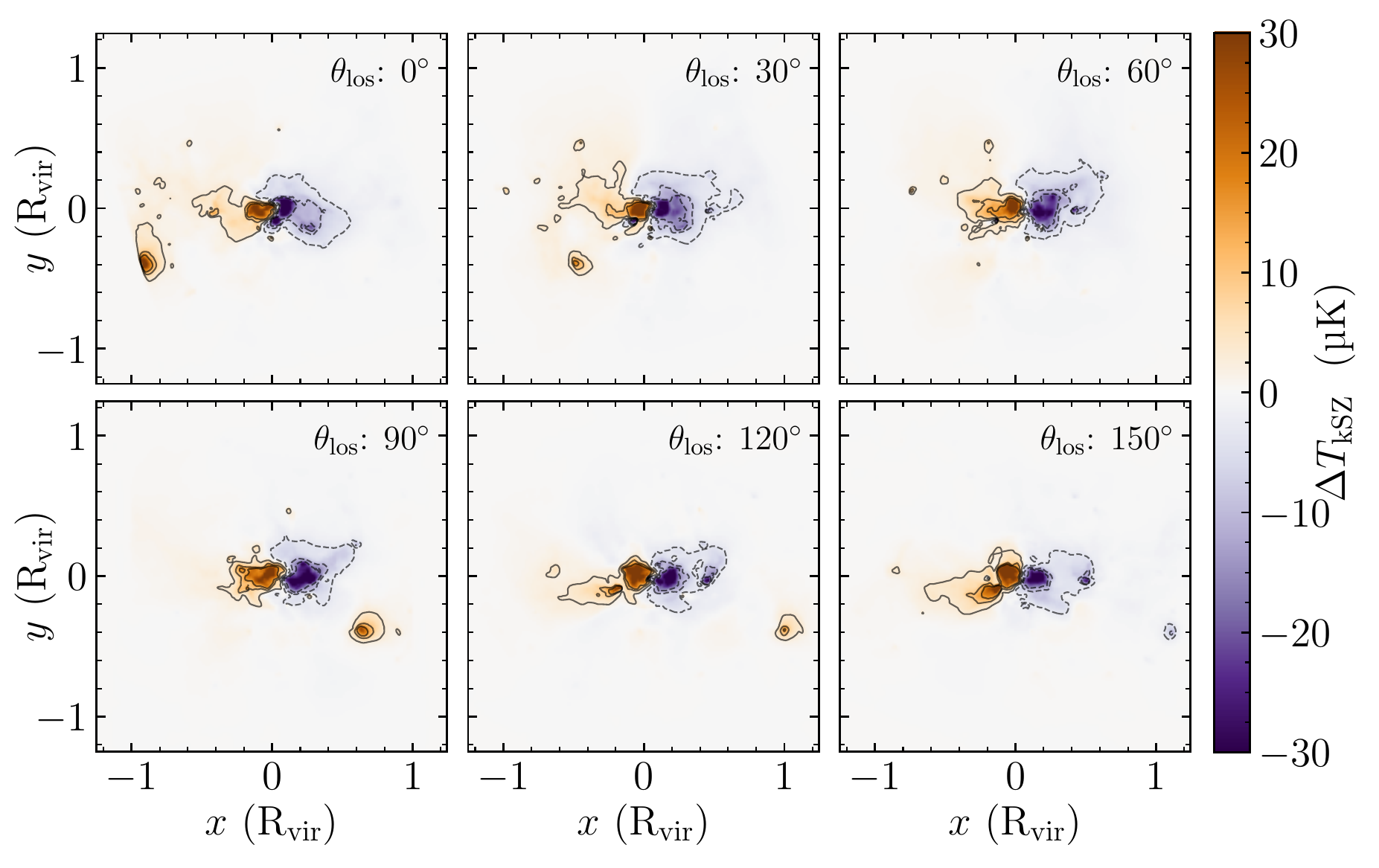}}
	\subfloat[cl. 256]{\includegraphics[width=0.50\textwidth]{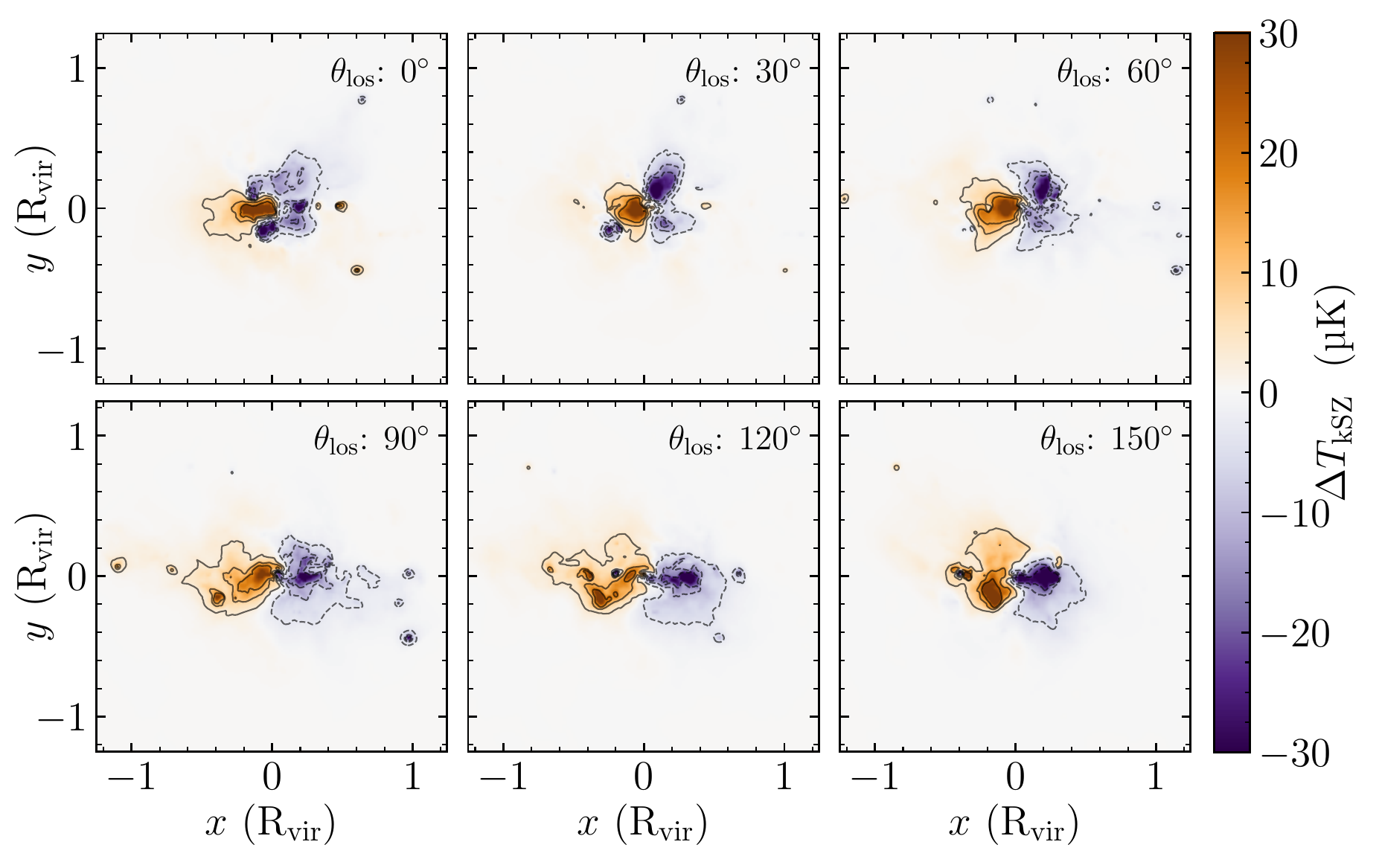}}
	\caption{\small Maps of the temperature shift produced by the kSZ effect for all the clusters in the sample, 
		obtained from different projections as described in the text, and smoothed at 20 arcsec.
		The angles of the corresponding lines of sight, taken on the plane orthogonal to the rotation axis, are specified
		on top of each map.
		Contours are plotted from -5$ \sigma $ to 5$ \sigma $, with dashed(solid) lines for negative(positive) values.
		The ranging values in the map have been set to $ \pm \SI{30}{\micro\K} $ for displaying purposes
		(see colour version of the figure in the online edition).}
	\label{fig:data_allos_allclus}
\end{figure*}

\begin{figure*}
	\centering
	\subfloat[cl. 46]{\includegraphics[width=0.50\textwidth]{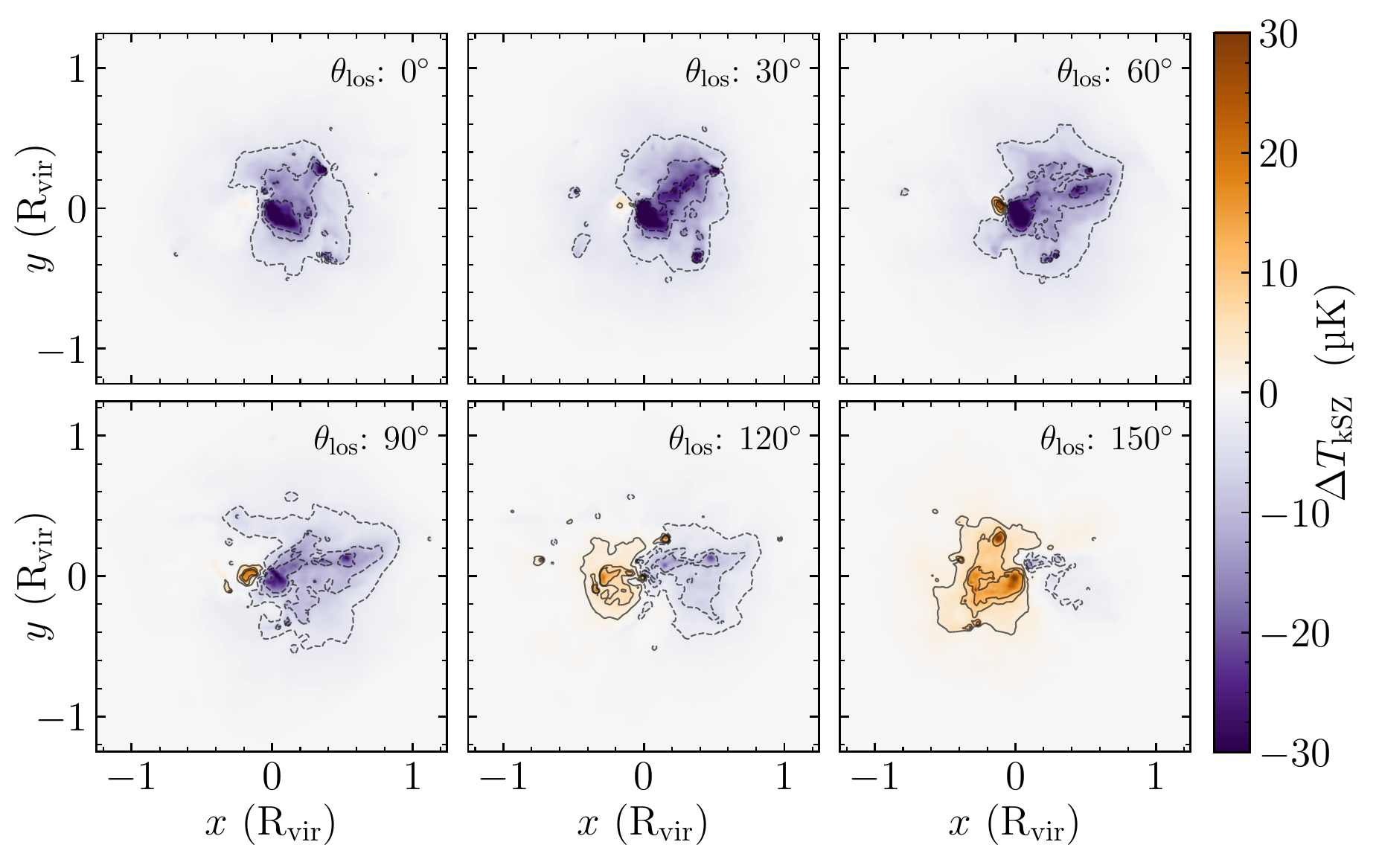}}
	\subfloat[cl. 93]{\includegraphics[width=0.50\textwidth]{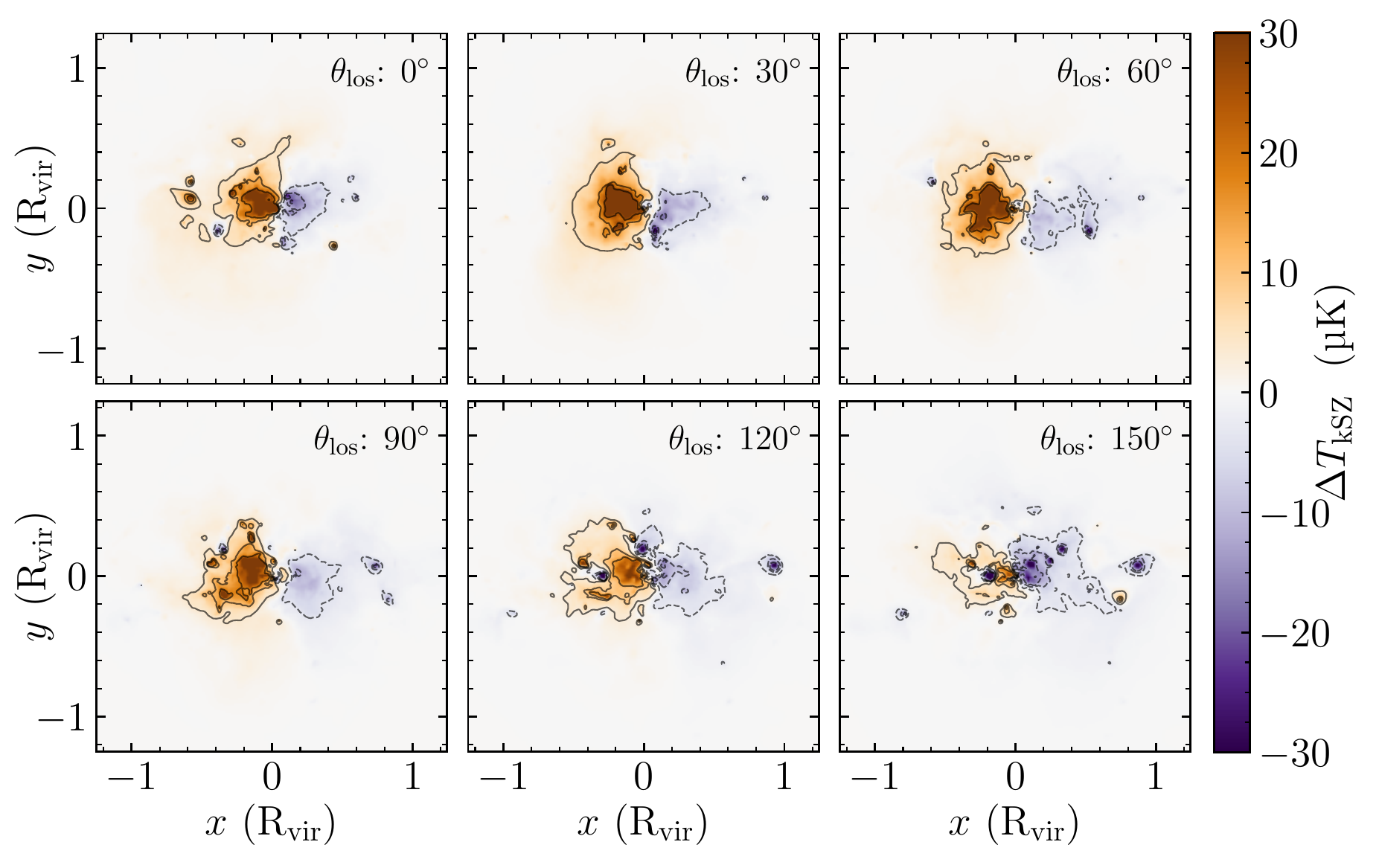}}\qquad
	\subfloat[cl. 98]{\includegraphics[width=0.50\textwidth]{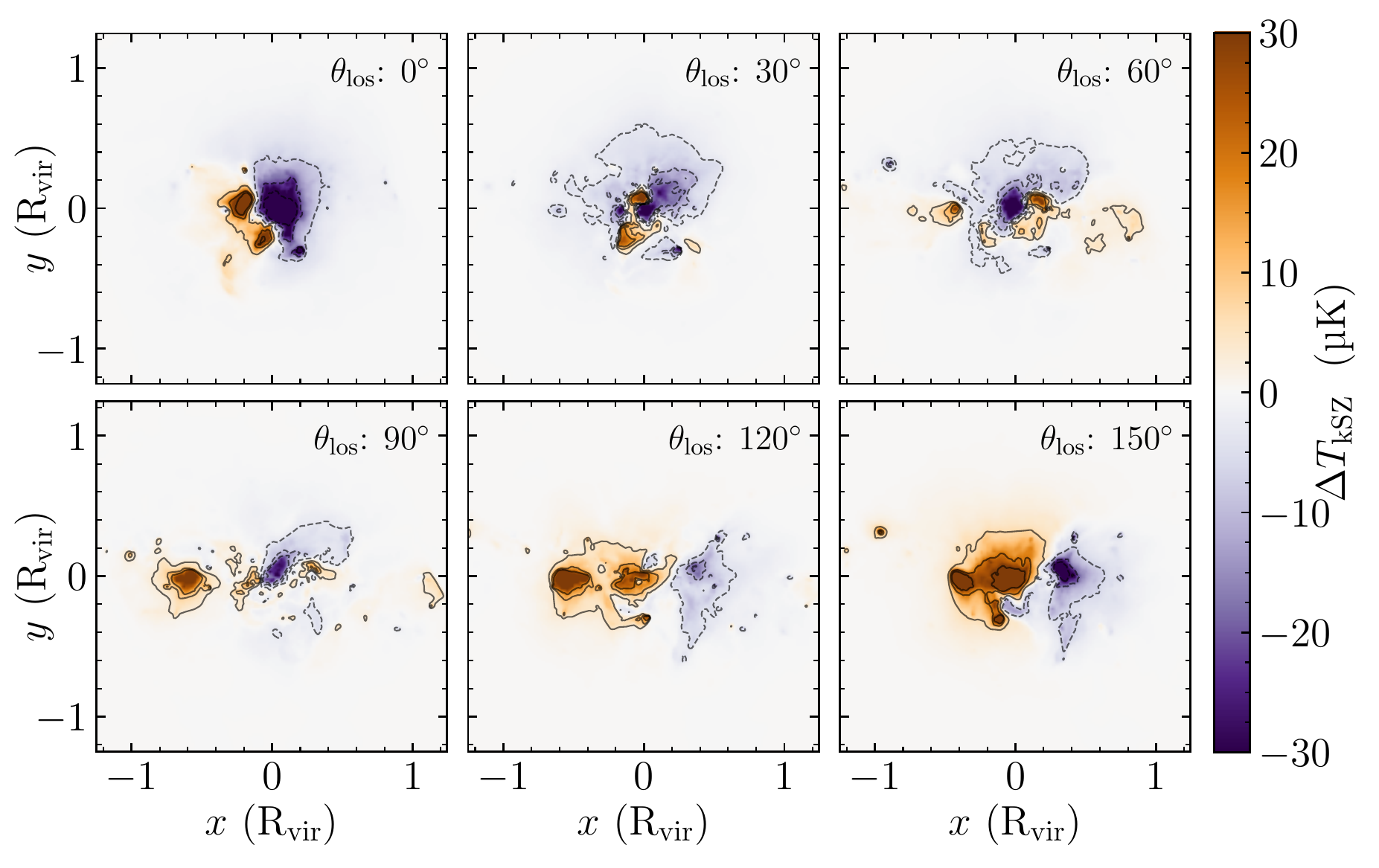}}
	\subfloat[cl. 103]{\includegraphics[width=0.50\textwidth]{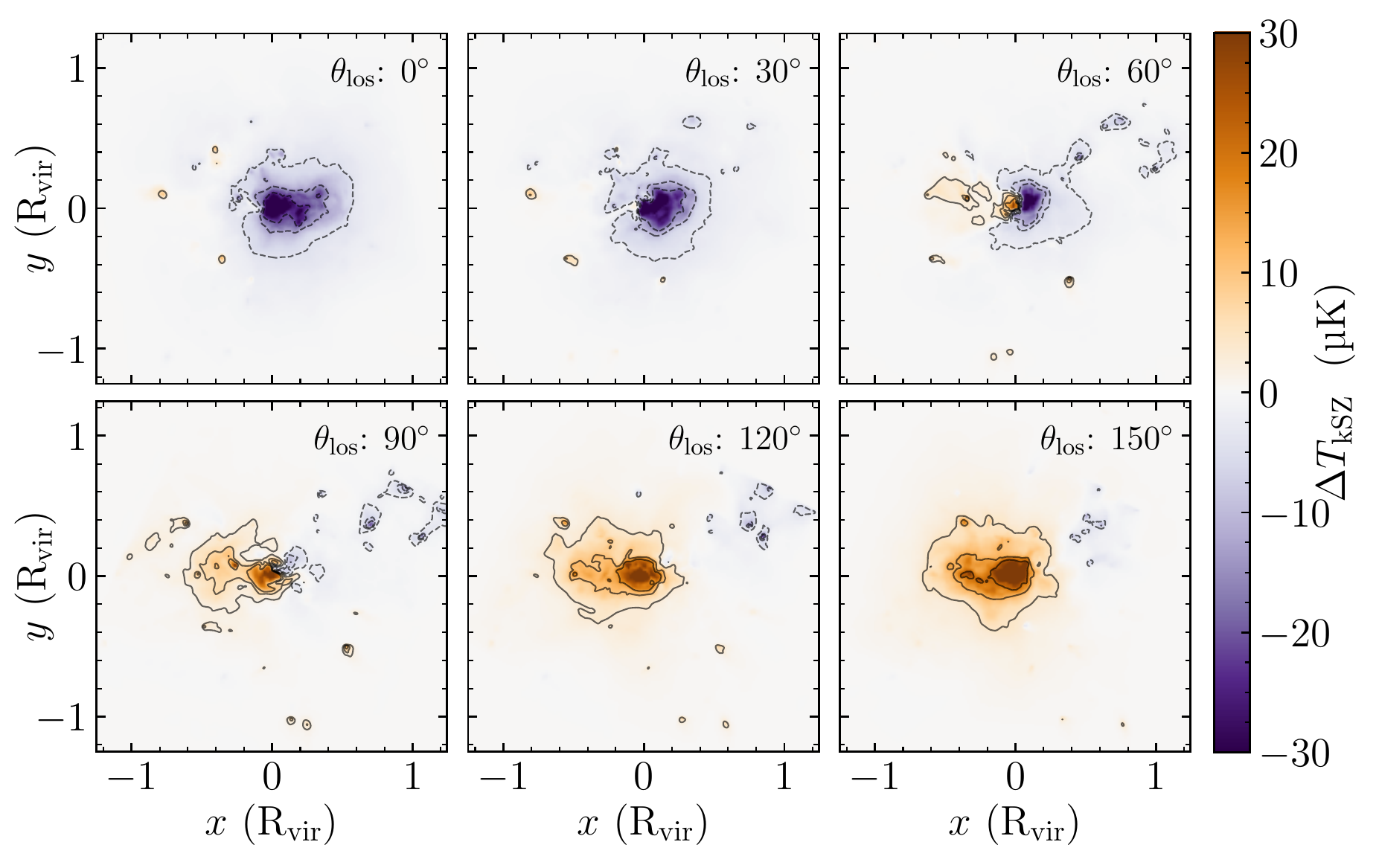}}\qquad
	\subfloat[cl. 205]{\includegraphics[width=0.50\textwidth]{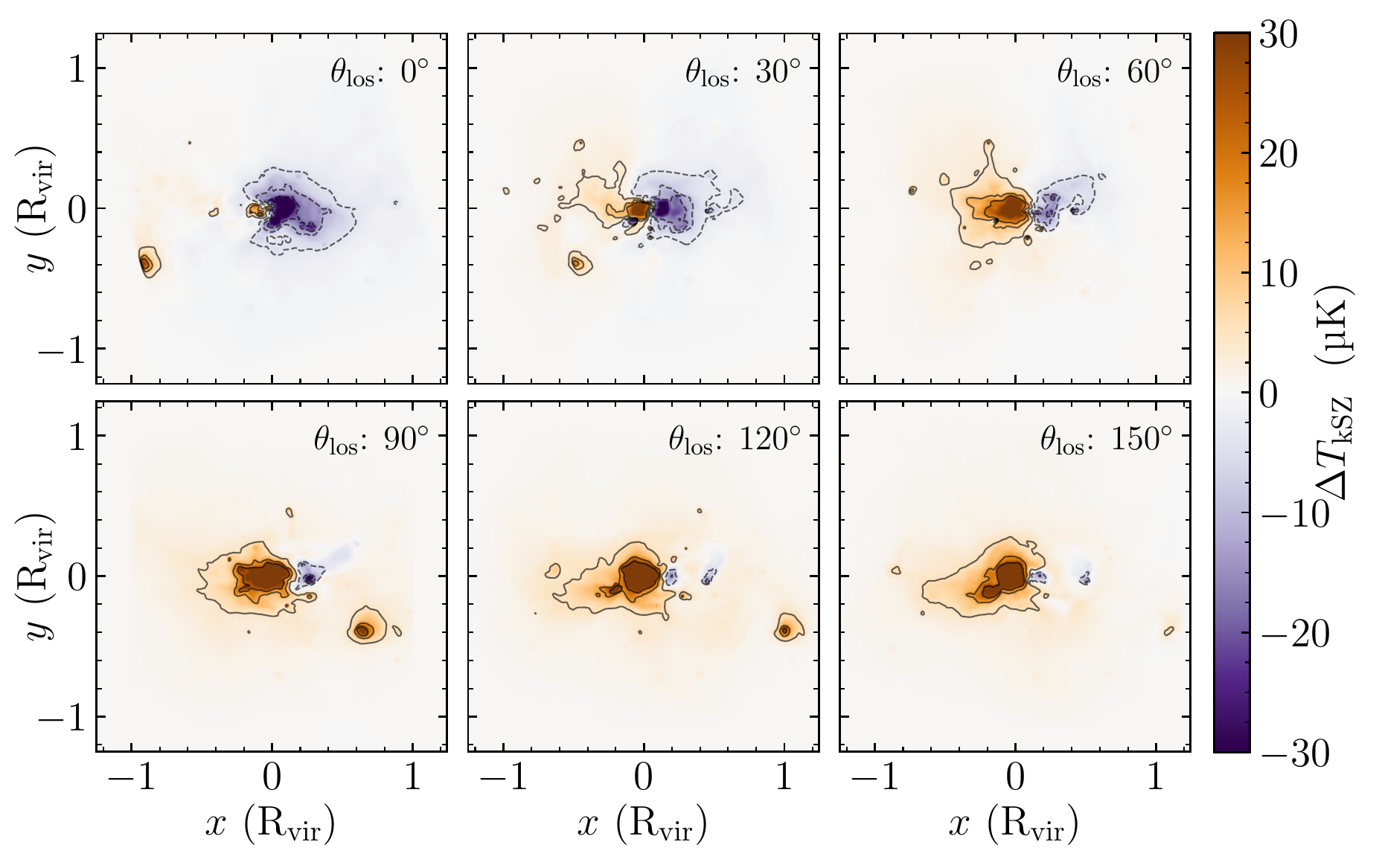}}
	\subfloat[cl. 256]{\includegraphics[width=0.50\textwidth]{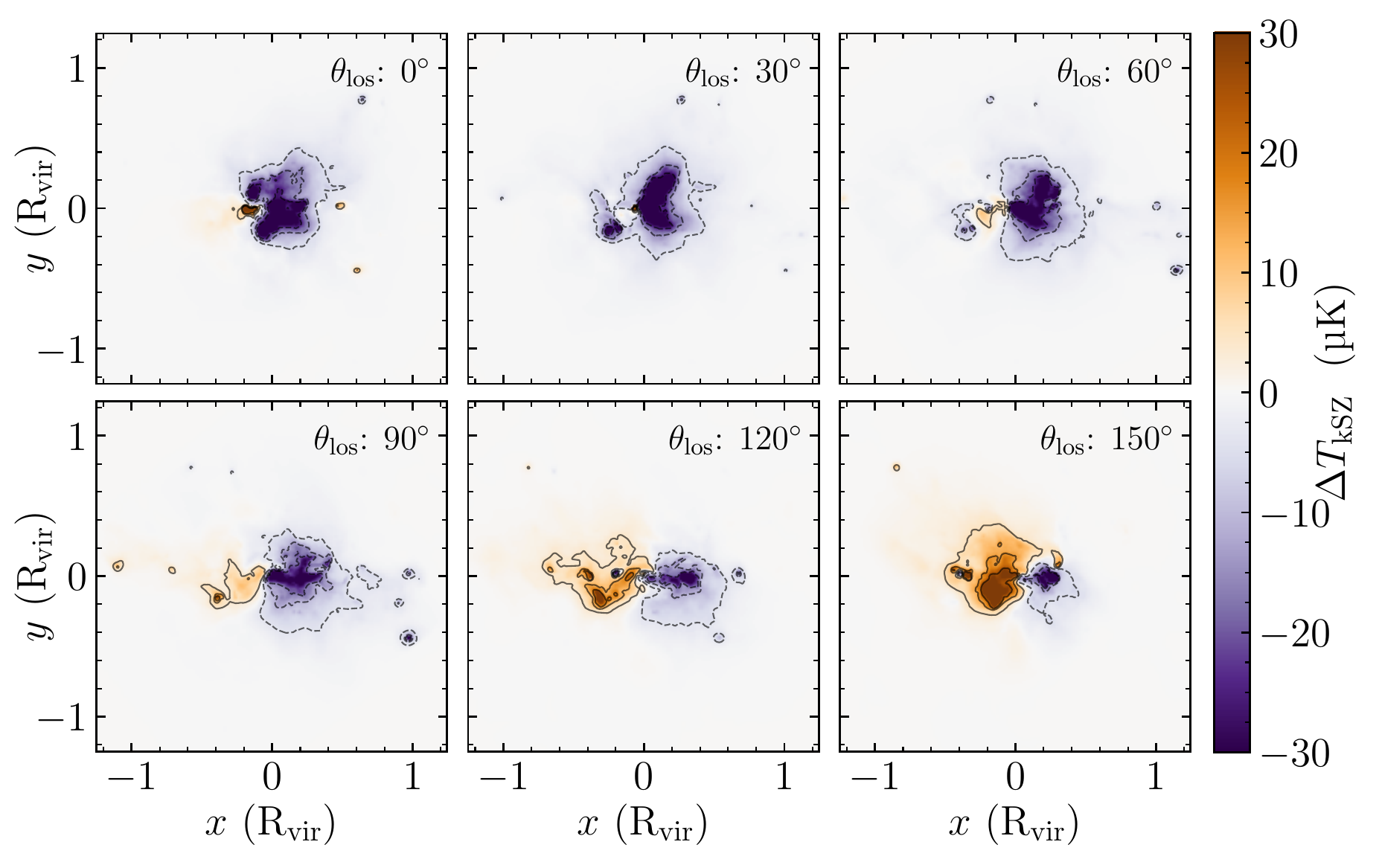}}
	\caption{\small Maps of the temperature shift produced by the kSZ effect accounting for the bulk motion
		for all the clusters in the sample,
		obtained from different projections as described in the text, and smoothed at 20 arcsec.
		The angles of the corresponding lines of sight, taken on the plane orthogonal to the rotation axis, are specified
		on top of each map.
		Contours are plotted from -5$ \sigma $ to 5$ \sigma $, with dashed(solid) lines for negative(positive) values.
		The ranging values in the map have been set to $ \pm \SI{30}{\micro\K} $ for displaying purposes
		(see colour version of the figure in the online edition).}
	\label{fig:data_allos_allclusbulk}
\end{figure*}

\begin{figure*}
	\centering
	\subfloat[cl.  46]{\includegraphics[width=0.4\textwidth]{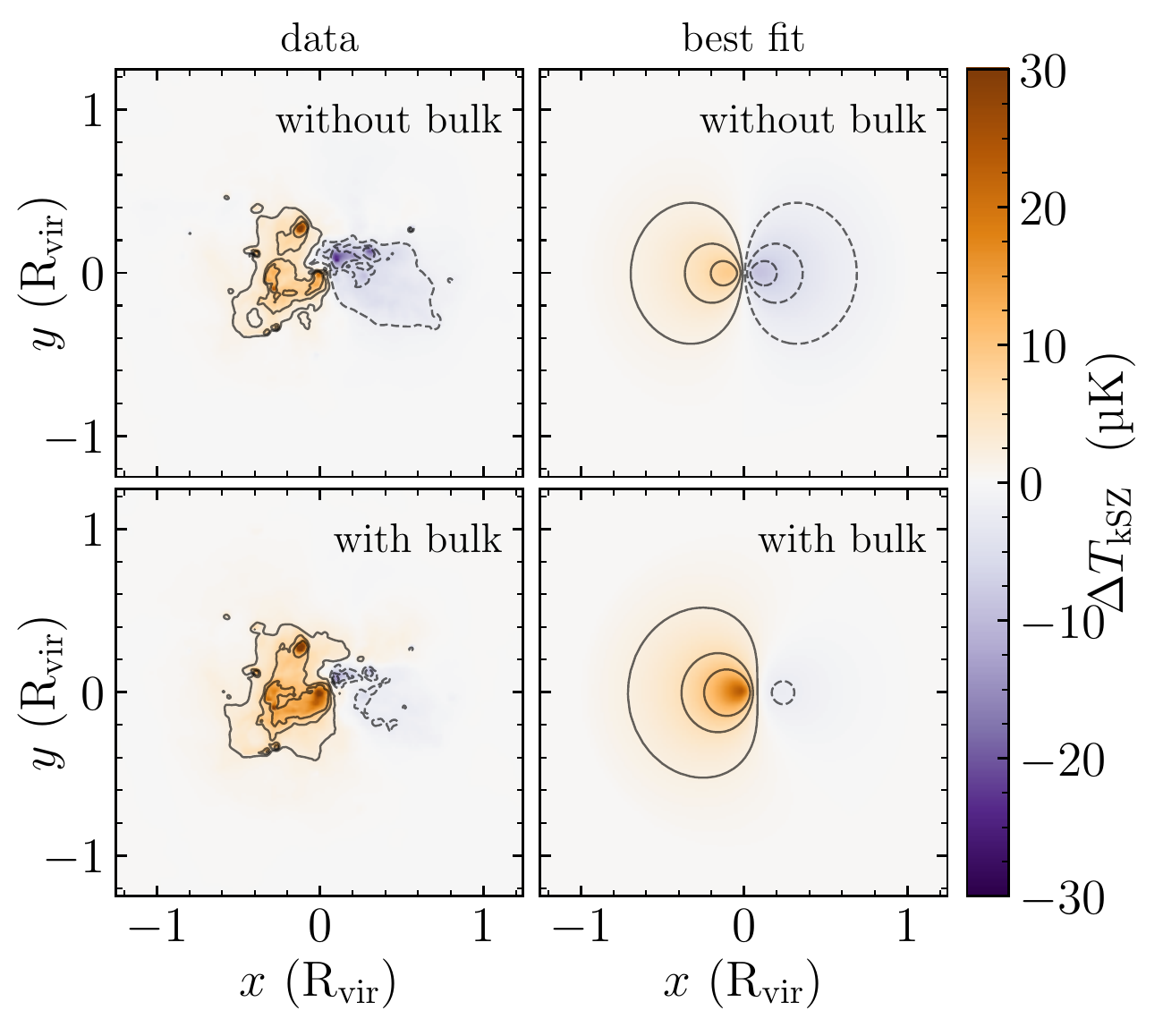}}\qquad
	\subfloat[cl.  93]{\includegraphics[width=0.4\textwidth]{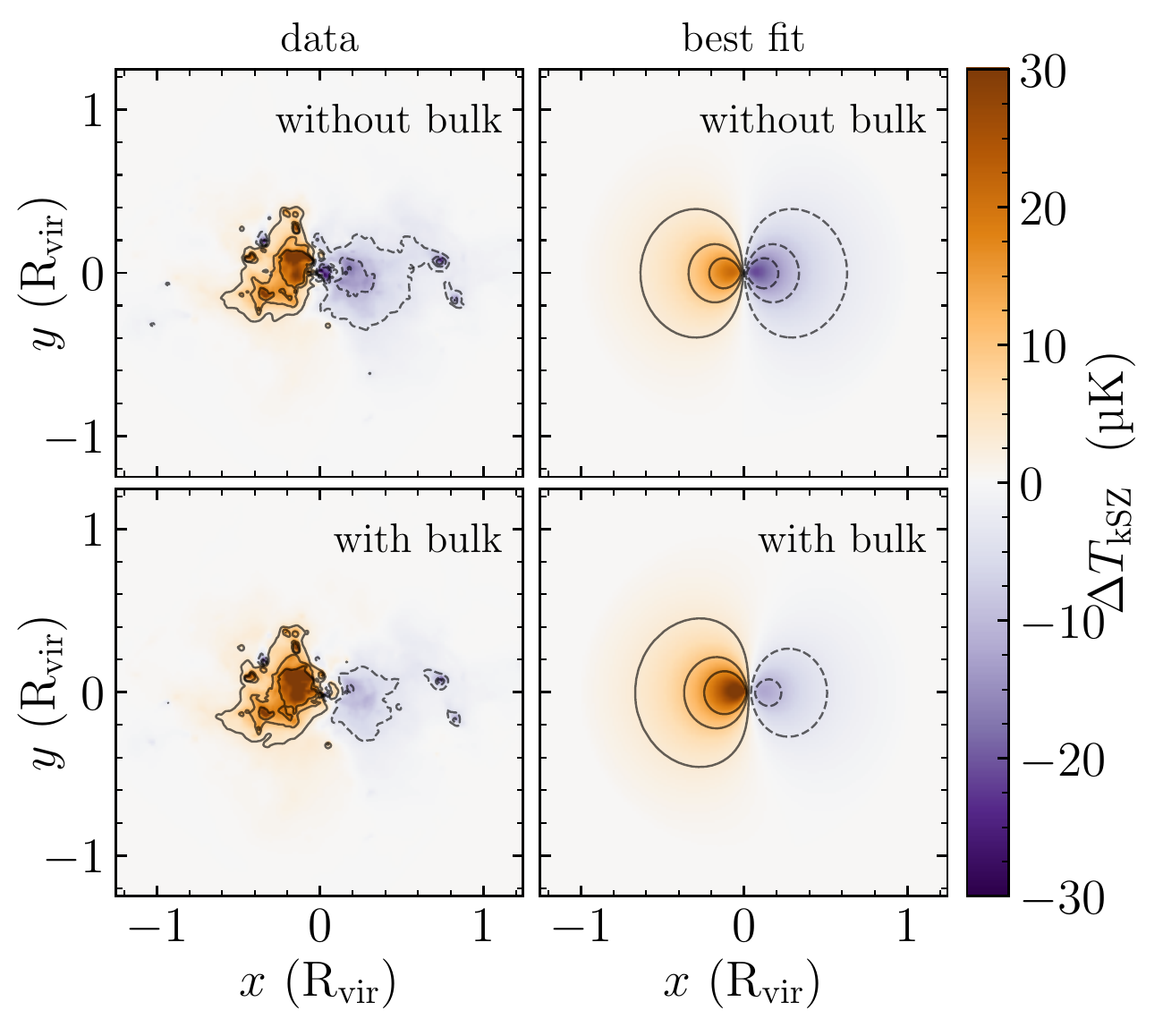}}\qquad
	\subfloat[cl.  98]{\includegraphics[width=0.4\textwidth]{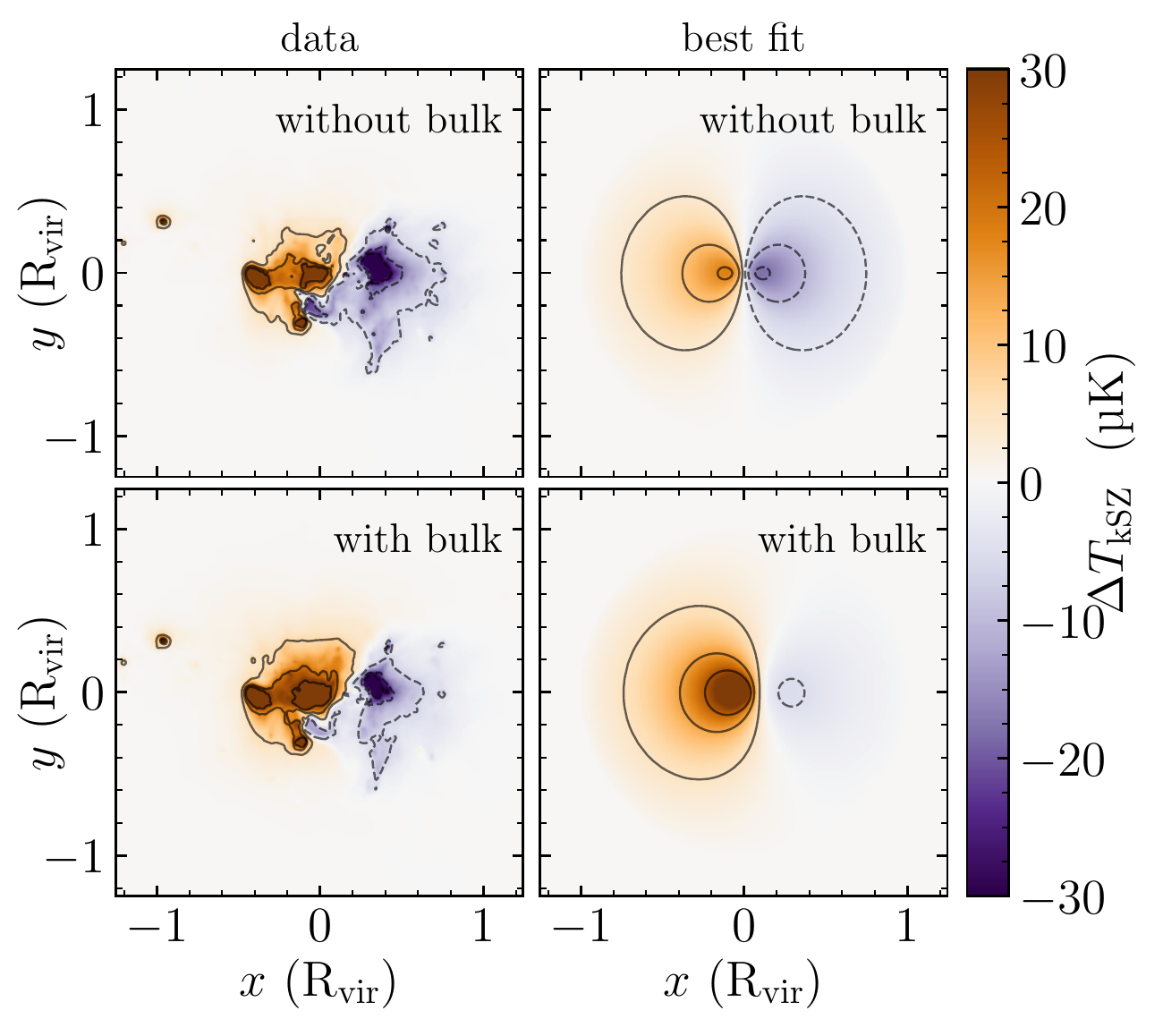}}\qquad
	\subfloat[cl. 103]{\includegraphics[width=0.4\textwidth]{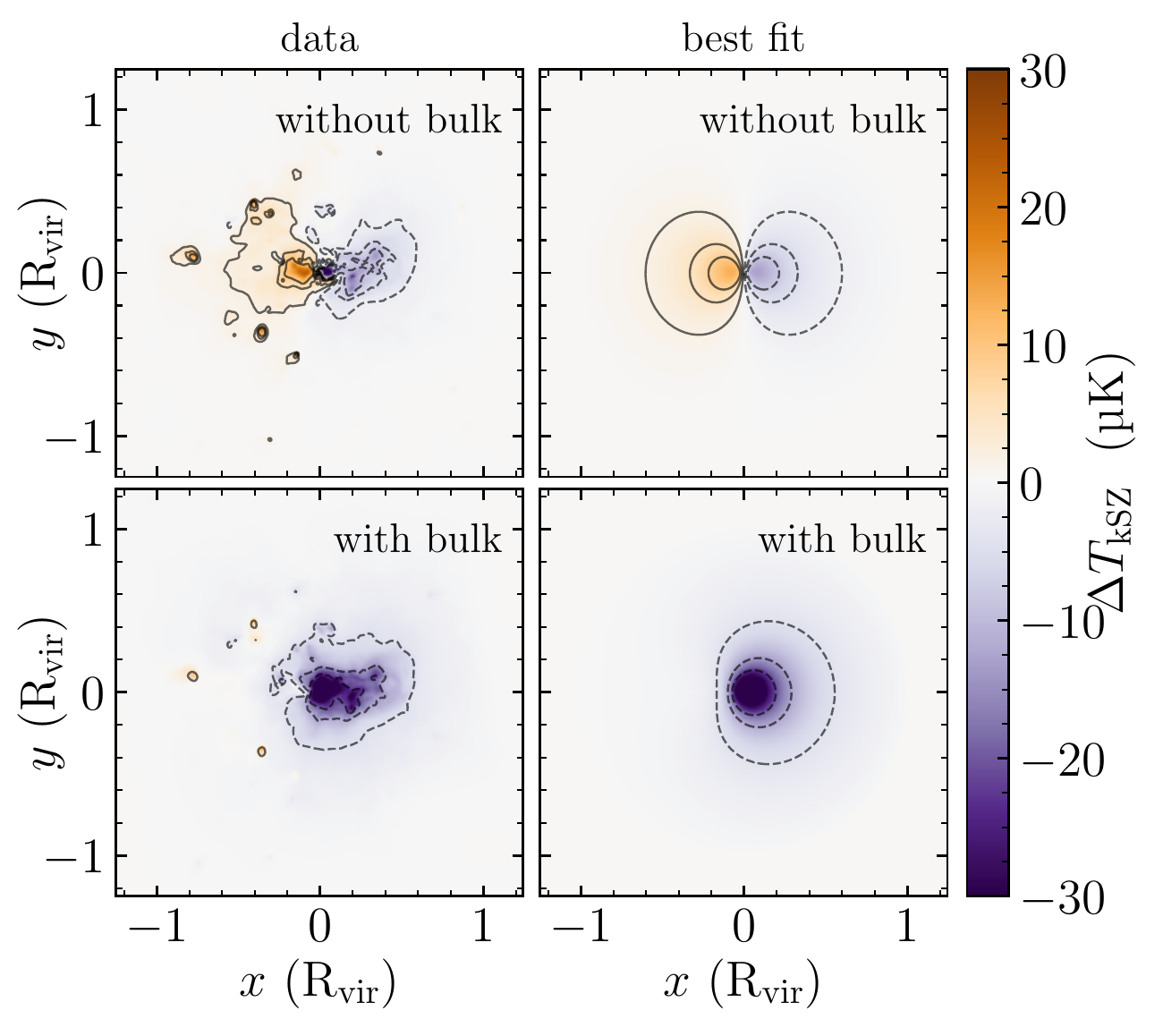}}\qquad
	\subfloat[cl. 205]{\includegraphics[width=0.4\textwidth]{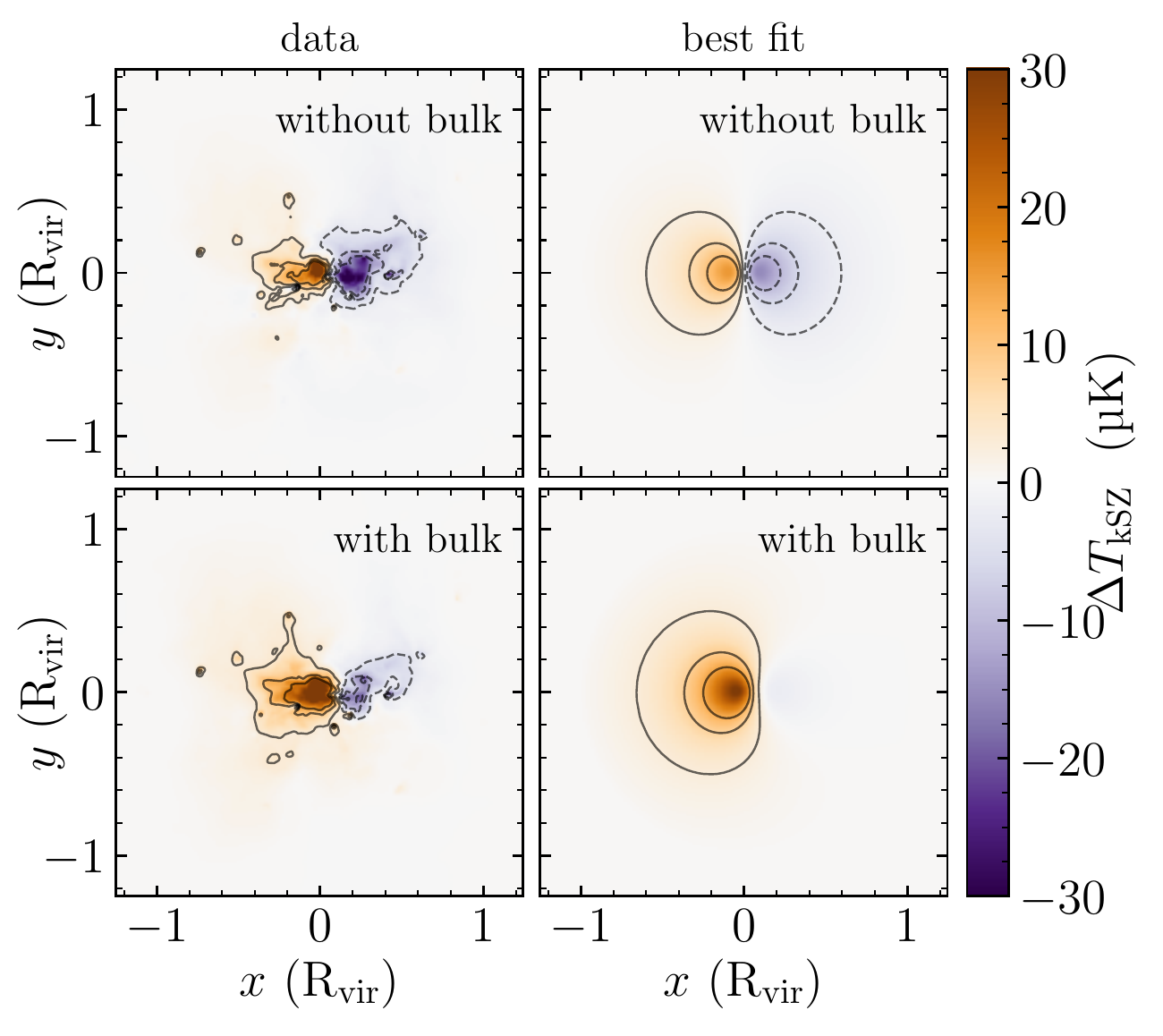}}\qquad
	\subfloat[cl. 256]{\includegraphics[width=0.4\textwidth]{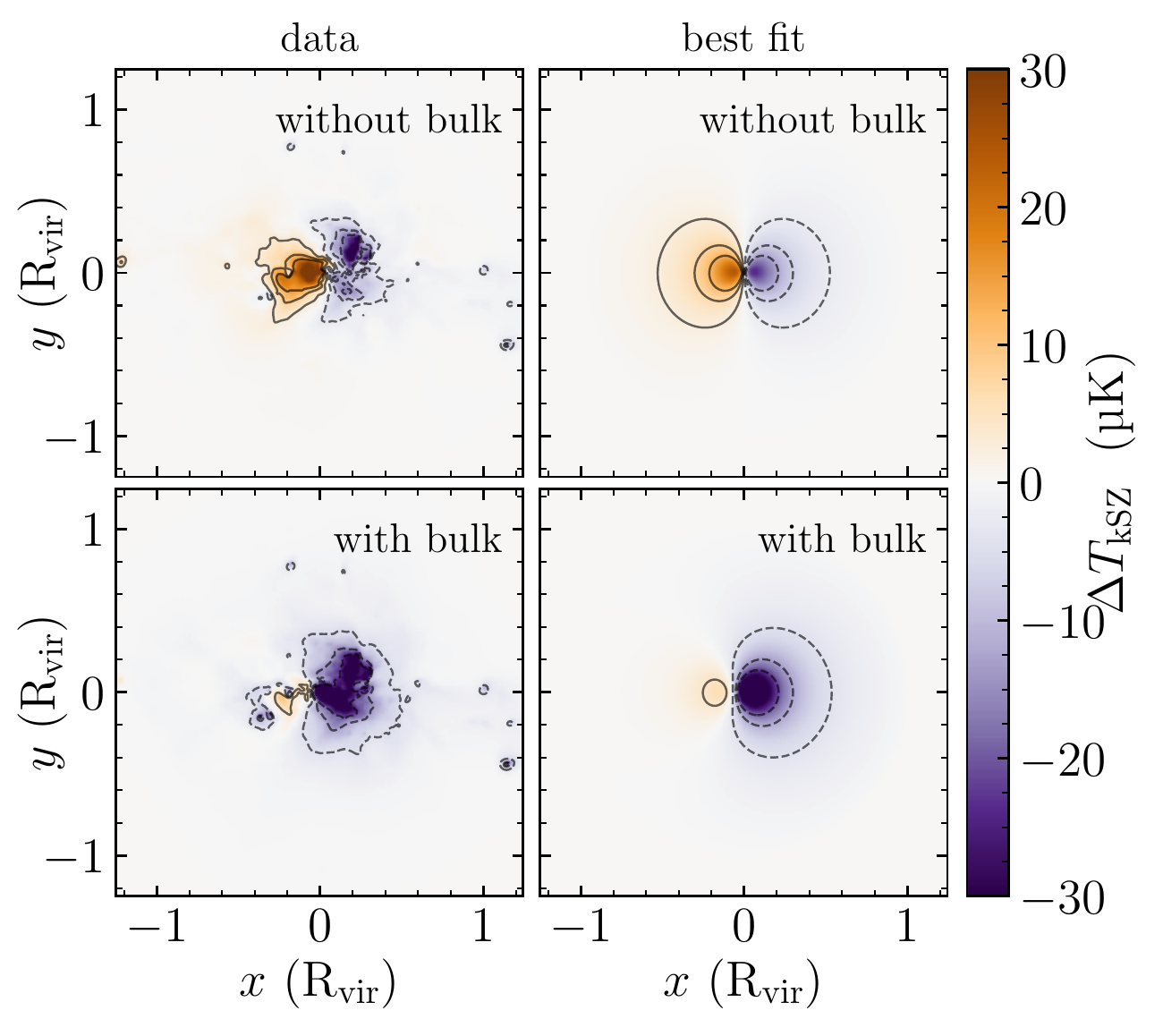}}
	\caption{\small Maps of the temperature shift produced by the kSZ effect at the best projection
		for all the clusters in the sample smoothed at 20 arcsec, with the corresponding best fit.
		Maps in the top and bottom panels refer to the case without and with the bulk velocity, respectively.
		Contours are plotted from -5$ \sigma $ to 5$ \sigma $, with dashed(solid) lines for negative(positive) values.
		The ranging values in the map have been set to $ \pm \SI{30}{\micro\K} $ for displaying purposes
		(see colour version of the figure in the online edition).}
	\label{fig:fits_bestlos_allclus}
\end{figure*}

We show in this section the kSZ temperature maps of all the six clusters in the sample.
The data maps, computed as detailed in section~\ref{sec:kszmaps} and smoothed at 20 arcsec, are shown in
Fig.~\ref{fig:data_allos_allclus} without the bulk motion. Fig.~\ref{fig:data_allos_allclusbulk} shows the same
maps without the subtraction of the bulk velocity.
The best fit maps together with the data maps are shown only for the best lines of sight in
Fig.~\ref{fig:fits_bestlos_allclus}, without (top panels) and with (bottom panels) the bulk motion.
\label{sec:appendix}
